\documentclass{sig-alternate-10pt}

\usepackage{fixltx2e}
\usepackage{amsmath}
\usepackage{float}
\usepackage{cite}
\usepackage{url}
\usepackage{booktabs} % For formal tables
\usepackage{epstopdf}

\usepackage{comment}
\usepackage[english]{babel}
\usepackage[utf8]{inputenc}
\usepackage{amsfonts}

\makeatletter
\newif\if@restonecol
\makeatother

\usepackage[linesnumbered,algoruled,boxed,lined]{algorithm2e}

\begin{document}
\title{Data Transfer Optimization Based on Offline Knowledge Discovery and Adaptive Real-time Sampling}

\numberofauthors{1}
\author{
\alignauthor
MD S Q Zulkar Nine, Kemal Guner, Ziyun Huang,\\ Xiangyu Wang, Jinhui Xu, and Tevfik Kosar\\
       \affaddr{Department of Computer Science and Engineering}\\
       \affaddr{University at Buffalo (SUNY)}\\
       \affaddr{Buffalo, New York 14260}\\
       \email{\{mdsqzulk,kemalgne,xiangyuw,jinhui,tkosar\}@buffalo.edu}
}    

\maketitle

\begin{abstract}
The amount of data moved over dedicated and non-dedicated network links increases much faster than the increase in the network capacity, but the current solutions fail to guarantee even the promised achievable transfer throughputs. In this paper, we propose a novel dynamic throughput optimization model based on mathematical modeling with offline knowledge discovery/analysis and adaptive online decision making. In offline analysis, we mine historical transfer logs to perform knowledge discovery about the transfer characteristics. Online phase uses the discovered knowledge from the offline analysis along with real-time investigation of the network condition to optimize the protocol parameters. As real-time investigation is expensive and provides partial knowledge about the current network status, our model uses historical knowledge about the network and data to reduce the real-time investigation overhead while ensuring near optimal throughput for each transfer. Our network and data agnostic solution is tested over different networks and achieved up to 93\% accuracy compared with the optimal achievable throughput possible on those networks.
\end{abstract}

\section{Introduction}
\label{sec:Introduction}
Applications in a variety of spaces --- scientific, industrial, and personal --- now generate
more data than ever before.
Large scientific experiments, such as 
high-energy physics simulations~\cite{CMS, ATLAS},
climate modeling~\cite{Climate, easterling2000climate},
environmental and coastal hazard prediction~\cite{Klein200335, carrara1999use},  
genomics~\cite{BLAST, morozova2008applications}, 
and astronomic surveys~\cite{loredo2007analyzing, eisenstein2011sdss}
generate data volumes reaching several Petabytes per year. Data collected from remote sensors and satellites, dynamic data-driven applications, digital libraries and preservations are also producing extremely large datasets for real-time or offline processing~\cite{Ceyhan07, Tummala07}. 
With the emergence of social media, video over IP, and more recently the trend for Internet of Things (IoT), we see a similar trend in the commercial applications as well, and 
it is estimated that, in 2017, more IP traffic will traverse global networks than all prior ``Internet years" combined. The global IP traffic is expected to reach an annual rate of 1.4 zettabytes, which corresponds to nearly 1 billion DVDs of data transfer per day for the entire year~\cite{Cisco_2016}.
 
As data becomes more abundant and data resources become more heterogenous, 
accessing, sharing and disseminating these data sets become a bigger challenge.
Managed file transfer (MFT) services such as Globus~\cite{globusonline}, PhEDEx~\cite{egeland2010phedex}, Mover.IO~\cite{moverio}, and B2SHARE~\cite{ardestani2015b2share} have allowed users to easily move their data, but these services still rely on the users providing specific details to control this process, and they suffer from inefficient utilization of the available network bandwidth and far-from-optimal end-to-end data transfer rates.
There is substantial empirical evidence suggesting that performance directly impacts revenue. As two well known examples for this, Google reported 20\% revenue loss due to a specific experiment that increased the time to display search results by as little as 500 milliseconds; and Amazon reported a 1\% sales decrease for an additional delay of as little as 100 milliseconds~\cite{Kohavi_2007}. 

End-to-end data transfer performance can be significantly improved by tuning the application-layer transfer protocol parameters (such as pipelining, parallelism, concurrency, and TCP buffer size). Sub-optimal choice of these parameters can lead to under-utilization of the network or may introduce congestion which would lead to queuing delay, packet loss, end-system over-utilization, extra power consumption, and other factors. It is hard for the end users to decide on optimal levels of these parameters statically, since static setting of these parameters might prove sub-optimal due to the dynamic nature of the network which is very common in a shared environment. 

In this paper, we propose a novel two-phase dynamic end-to-end data transfer throughput optimization model based on mathematical modeling with offline knowledge discovery/analysis and adaptive online decision making. During the  offline analysis phase, we mine historical transfer logs to perform knowledge discovery about the transfer characteristics. During the online phase, we use the discovered knowledge from the offline analysis along with real-time investigation of the network condition to optimize the protocol parameters. As real-time investigation is expensive and provides partial knowledge about the current network status, our model uses historical knowledge about the network and data to reduce the real-time investigation overhead while ensuring near optimal throughput for each transfer. We have tested our network and data agnostic solution over different networks and observed up to 93\% accuracy compared with the optimal achievable throughput possible on those networks. Extensive experimentation and comparison with best known existing solutions in this area revealed that our model outperforms existing solutions in terms of accuracy, convergence speed, and achieved end-to-end data transfer throughput. 

 In summary, the contributions of this paper include:
 \begin{enumerate}
  \item Instead of performing solely mathematical optimization on the fly, we use historical log analysis to construct the throughput surface for different parameters and external loads. We pre-compute the optimal solution for each during the offline analysis phase. 
  \item We construct all possible throughput surfaces in the historical data using cubic spline interpolation, and create a  probabilistic confidence region with Gaussian distribution to encompass each surface.
  \item Real production level data transfer logs are used in our experiment and to compute static parameter settings for different types of transfers.
  \item In real time, adaptive sampling technique is used over the pre-computed throughput surfaces to provide faster convergence towards maximally achievable throughput.
  \item We provide highly accurate prediction of the optimal transfer parameter combinations with minimal sampling overhead. 
\end{enumerate}

The rest of the paper is organized as follows: Section
II presents the problem formulation; Section III discusses our proposed model;
Section IV presents the evaluation of our model; Section
V describes the related work in this field; and Section VI
concludes the paper with a discussion on the future work.

\section{Problem Formulation}
\label{sec:Problem Formulation}
Application level data transfer parameters can have different impact on files with different sizes and the number of files in the dataset. \textbf{Concurrency}, $cc$ controls the number of server processes which can transfer different files concurrently~\cite{kosar04, Thesis_2005, Kosar09, JGrid_2012}. It can accelerate the transfer throughput when a large number of files needs to be transferred. \textbf{Parallelism}, $p$ is the number of data connections that each server process can open to transfer the different portion of the same file in parallel~\cite{R_Sivakumar00, R_Lee01, R_Balak98, R_Hacker05, R_Eggert00, R_Karrer06, R_Lu05, DADC_2008, NDM_2011}. It can be a good option for large or medium files. Therefore, the number of parallel data streams is $(cc\times p)$. \textbf{Pipelining}, $pp$ is useful for small file transfers~\cite{TCP_Pipeline, farkas2002, Cluster_2015, TCC_2016}. It eliminates the delay imposed by the acknowledgment of the previous file before starting the next file transfer. For high latency wide-area networks, this delay might prove highly sub-optimal. 

Given a source endpoint $e_s$ and destination endpoint $e_d$ with a link bandwidth $b$, round trip time $rtt$, and a dataset of size $f_{all}$, average file size $f_{avg}$, number of files $n$ and set of protocol parameters $\theta=\{cc, p, pp\}$, the throughput, $th$ optimization problem can be defined as:

\begin{equation}
\underset{ \{cc,p,pp\} }{\mathrm{argmax}} \int_{t_s}^{t_f} th(e_s,e_d,b,rtt,f_{avg}, n,cc,p,pp,l_{ctd},l_{ext})
\end{equation}

Where $l_{ctd}$ and $l_{ext}$ are the external load from other transfers. As we are optimizing throughput function in a shared environment, other concurrent transfers can affect the behavior of achievable throughput. We can account the incoming and outgoing transfers happening from the source and destination nodes. Our historical logs contain information of such transfers. We define those contending transfers as $t_{ctd}$. There might exist other transfers with little-known information. We define those transfers as external load $t_{ext}$. 

We have made couple of assumption to define our model. Those are expressed here. 
\newtheorem{assumption}{Assumption}
\begin{assumption}\label{as:1}
  Competing Transfers can achieve aggregate throughput, $T = \sum_{i=1}^{N} th_i$, where $N$ is the number of TCP streams for all competing transfers. 
\end{assumption}

\begin{assumption}\label{as:2}
  After explaining away the effect of known competing transfers, the fluctuation on transfer behavior depends on the intensity of the external load $t_{ext}$.  
\end{assumption}

\begin{assumption}\label{as:3}
Maximum achievable throughput can be bounded by bandwidth, disk read or disk write bottleneck.   
\end{assumption}

Given disk read speed, $v_{read}$ and disk write speed $v_{write}$ and the link bandwidth $b$, the maximally achievable throughput, 

\begin{assumption}\label{as:4}
Our model is a network protocol optimization and it is underlying file system agnostic. It can provide superior performance with a parallel file system. 
\end{assumption}

Performance degradation due to hardware misconfiguration, storage access delay, intermediate network devices bottleneck, etc. could limit the achievable throughput. Eliminating such bottlenecks might increase the limit of achievable throughput.

\section{Proposed Model}
\label{sec:Proposed Model}
Our model consists of two phases: \textit{(i)} Offline knowledge discovery \textit{(ii)} Online Adaptive sampling. Offline Analysis module is an additive model. That means when new logs are generated for a certain period of time, we do not need to combine it with previous logs and perform analysis on whole log \textit{(old log + new log)}. Users do not need to perform offline analysis during each transfer. Data transfer logs can be collected for a certain period of time and then offline analysis can be performed on those new logs. For services like Globus, historical logs can be analyzed by a dedicated server and results can be shared by the users. When a user starts data transfer process, system initiates online adaptive sampling. Adaptive sampling queries the results of offline analysis which can be answered in constant time. Then adaptive sampling guided by offline analysis provides faster convergence towards near optimal throughput.  

\subsection{Offline Analysis}
\label{subsec:Offlineanalysis}
Offline analysis collects useful information from the historical logs so that those information can be used by online module to converge faster. Offline analysis consist of five phases - \textit{(i)} Clustering logs in hierarchy, \textit{(ii)} Surface construction, \textit{(iii)} Find maximal parameter setting, \textit{(iv)} Accounting for known contending transfers, \textit{(v)} Identify suitable sampling regions. 

Historical logs contain information about the verity of transfers performed by the users. Therefore, a natural approach would be cluster the logs based on different matrices. Assuming that we have historical log, $L$ of $n_{log}$ log entries, We can define our clustering problem as $(L,m)$, where $m$ is the number of target clusters. The clusters of the historical logs are $C = \{ C_{1},...,C_{m}\}$, where $\{n_{1},...,n_{m}\}$ denote the sizes of the corresponding clusters. We consider a pair-wise distance function, $\it{d}(x,x^\prime)$ where $x,x^\prime \in L$. We have tested clustering algorithm for different pair-wise distance functions. For clustering, we have tested two well-known approaches - (1) K-means++~\cite{kmeans++}, (2) Hierarchical Agglomerative  Clustering (HAC) with Unweighted Pair Group Method with Arithmetic Mean (UPGMA)~\cite{upgma_clustering}. K-means clustering algorithm suffers from initial centroid selection. Wrong initialization could lead to wrong clustering decision. However, K-means++ provides theoretical guarantee to find a solution that is $O(log\ m)$ competitive to the optimal k-means solution. For HAC, distance between two clusters $C_{i}$ and $C_{j}$ is defined as $D(C_{i},C_{j})$ in Equation (\ref{eq:UPGMAdistance}):

\begin{equation}
\label{eq:UPGMAdistance}
D(C_{i},C_{j}) = d(c_i,c_j) = \sqrt[]{(c_i - c_j)^{2}}
\end{equation}

Where $c_i$ and $c_j$ are the corresponding cluster centroid of $C_i$ and $C_j$. HAC computes proximity matrix for the initial clusters. Then it combines two clusters with minimum distance $D(C_i,C_j)$. Then it updates the rows and columns of proximity matrix with new clusters and fills out the matrix with new $D(C_i,C_j)$. This process is repeated until all clusters are merged into a single cluster. 

Clustering accuracy depends on the appropriate number of clusters $k$. In this work, we have used Calinski and Harabasz index(CH index) to identify the appropriate number of clusters. CH index can be computed as:
\begin{equation}
CH(m) = \dfrac{\Phi_{inter}(m)/(m-1)}{\Phi_{inter}(m)/(n-m)}
\end{equation}
Where $\Phi_{inter}$ is the Between-cluster variation and $\Phi_{intra}$ is the Within-cluster variation. Both can be defined as sum of Euclidean distance as bellow:

\begin{equation}
 \Phi_{inter}(m) = \sum^{m}_{i=1} \sum_{x \in c_i} (x - c_i)^2
\end{equation}

\begin{equation}
 \Phi_{intra}(m) = \sum^{m}_{i=1} n_k(\bar{C_k} - \bar{x})^2
\end{equation}
Where $\bar{C_k}$ is the mean of points in cluster $k$ and $\bar{x}$ is the overall mean. Largest $CH(m)$ score is preferable. 

%\begin{comment}
%%%%%%%% Surface Construction sub sub section %%%%%%%%%%%%%
\begin{figure*}[t]
    \centering
        \includegraphics[keepaspectratio=true,width=55mm]{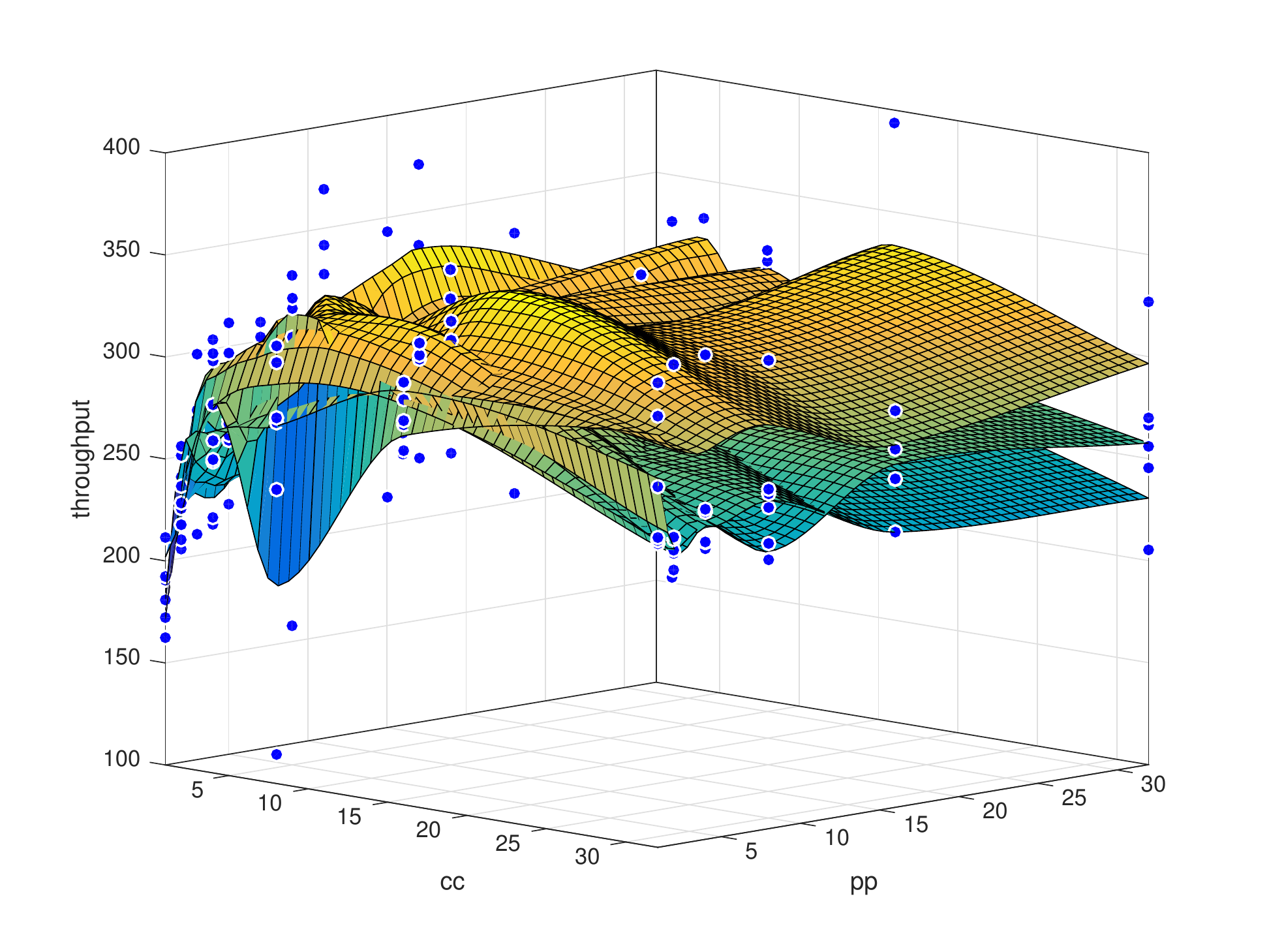}
        \includegraphics[keepaspectratio=true,width=55mm]{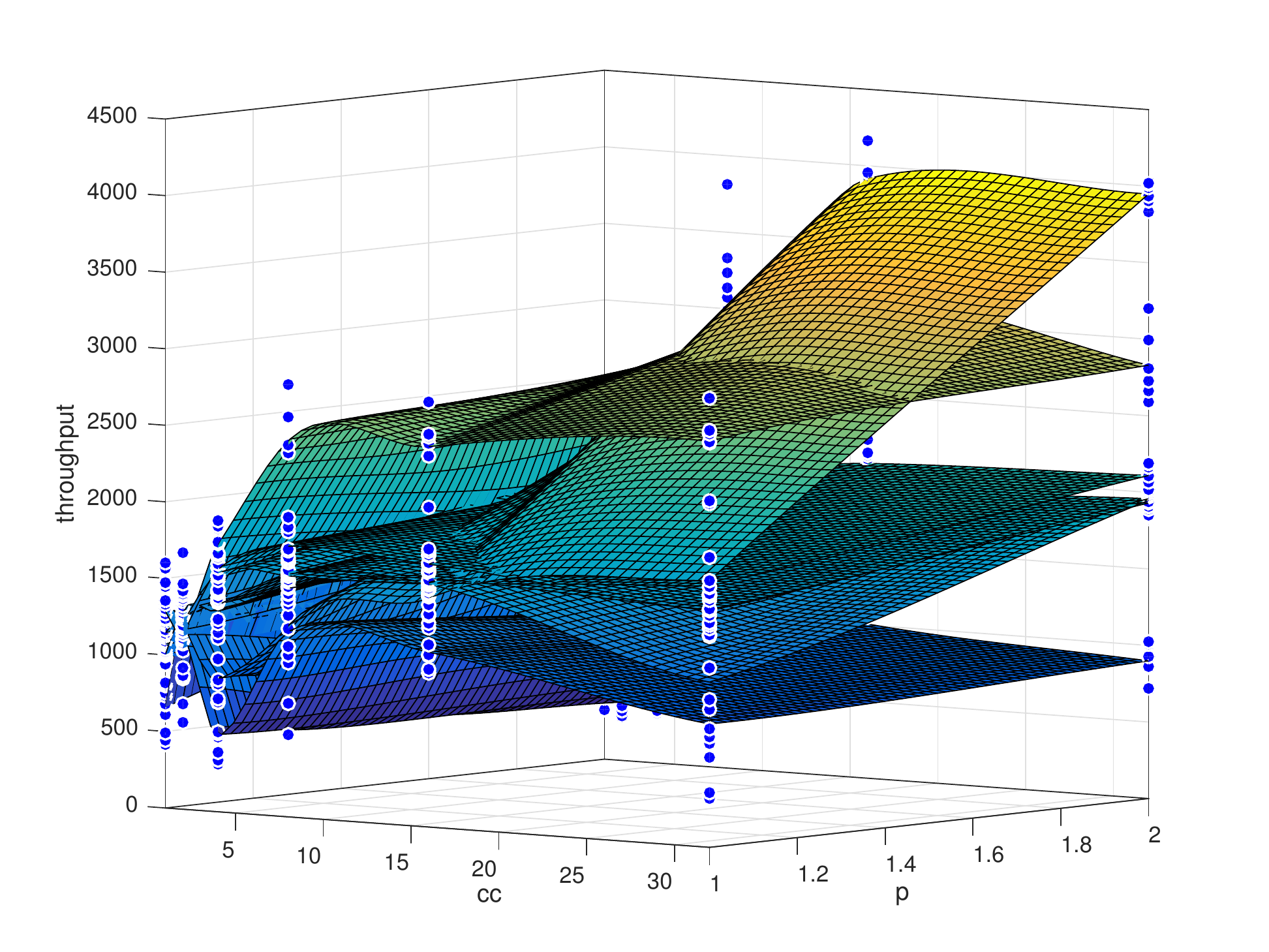}
        \includegraphics[keepaspectratio=true,width=55mm]{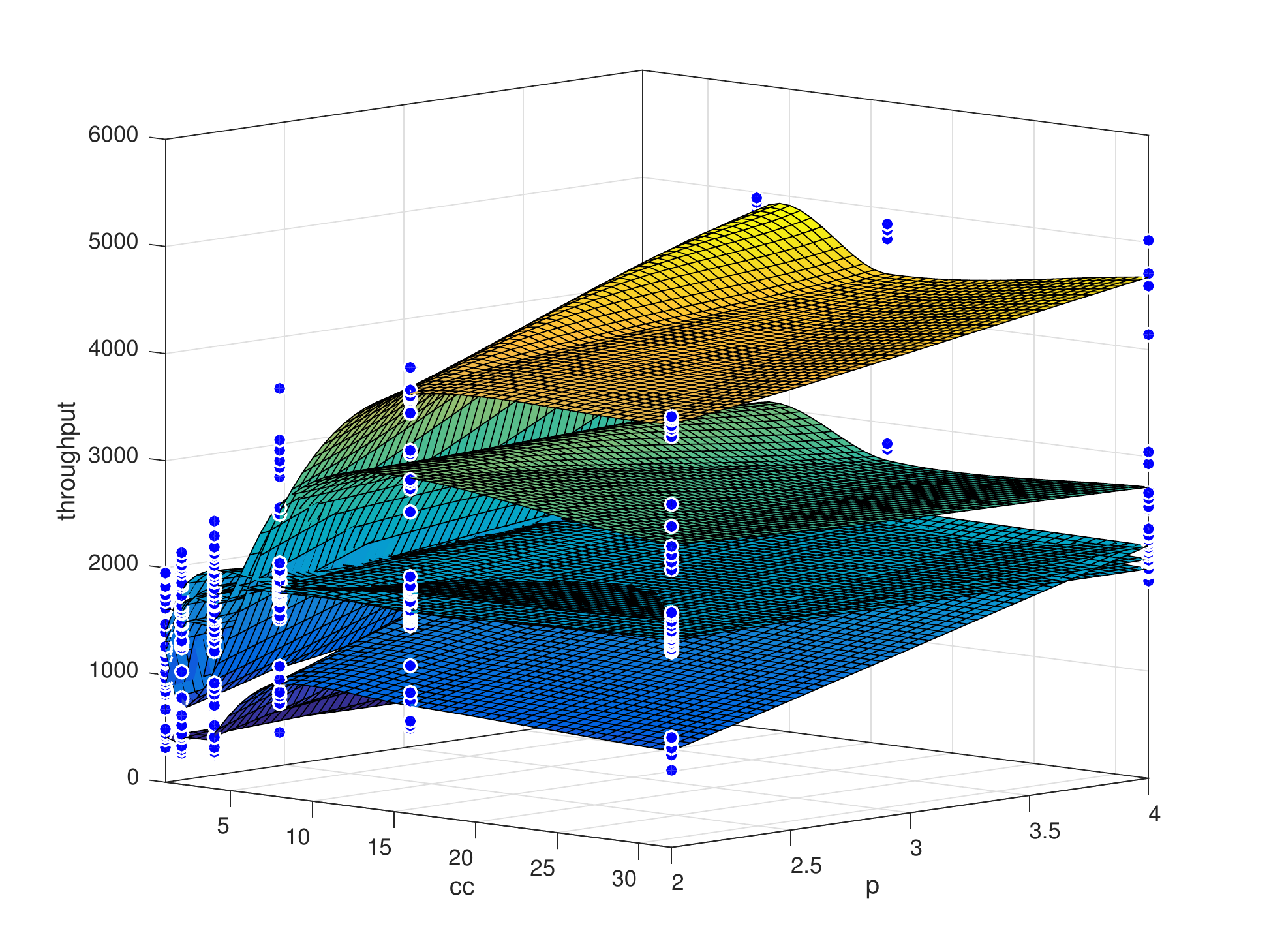}
     \caption{Piecewise cubic interpolation surface construction}
     \label{fig:3diinterpolation}
 \end{figure*}
%\end{comment}

%\begin{comment}
\begin{figure}[t]
	\begin{centering}
\includegraphics[keepaspectratio=true,angle=0, height= 70mm, width=130mm]{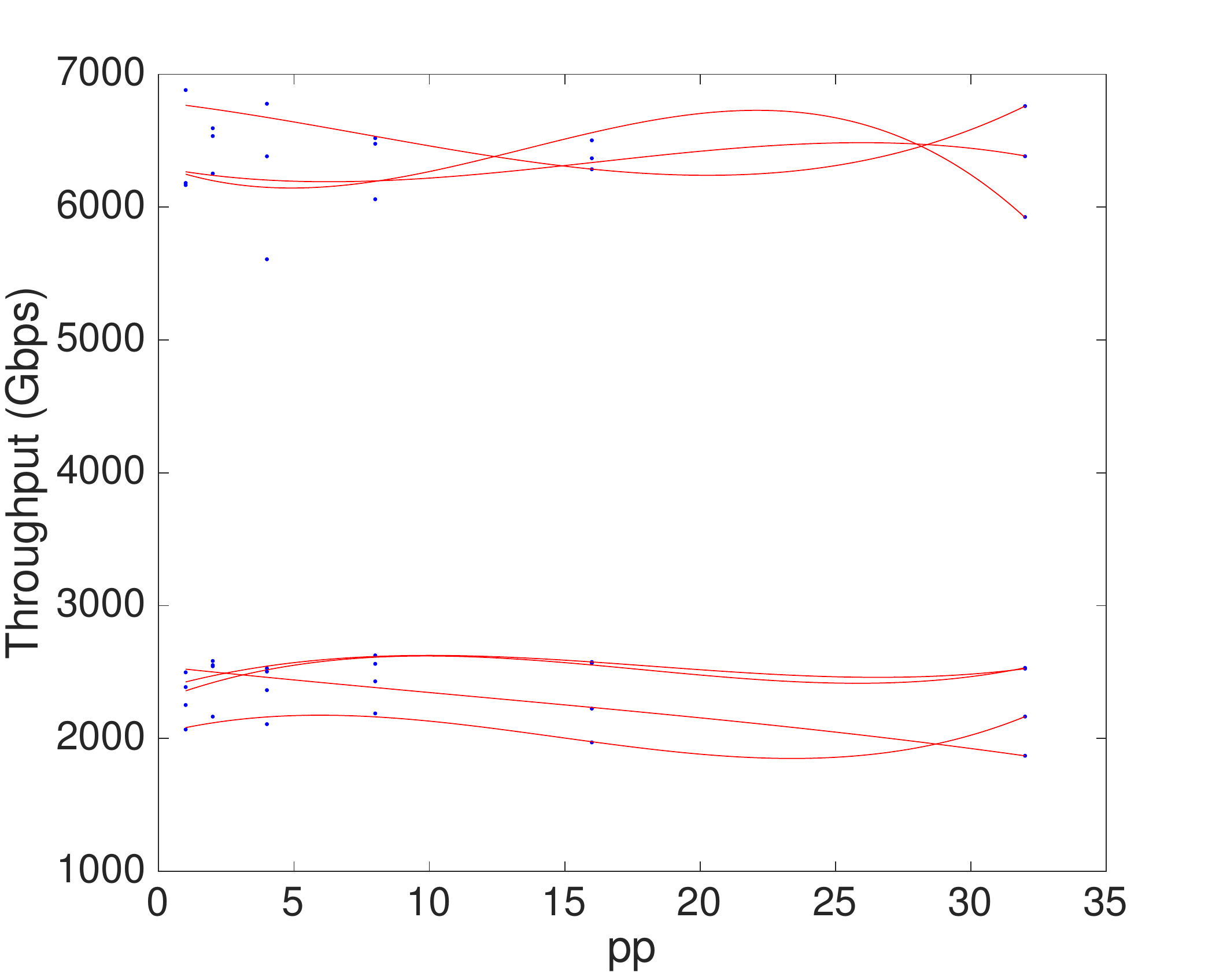}
	\end{centering}
    \vspace{-3mm}
\caption{Piecewise cubic interpolation of throughput over pipelining } \label{fig:ppCurve}
\end{figure}
%\end{comment}

\subsubsection{Surface construction}
\label{subsubsec:surfaceconstruction}
Achievable throughput for a given cluster $C_{i}$ can be modeled as a polynomial surface which depends on the protocol tuning parameters. We have tried three models to see how accurate those can capture the throughput behavior. The models are - (1) Quadratic regression, (2) Cubic regression and (3) Piecewise cubic interpolation. 

\textbf{\textit{(1) Quadratic regression:  }} A quadratic surface can be represented as: 
\begin{equation}
f(p,cc,pp) = t_{1}p^{2} + t_{2}cc^{2} + t_{3}pp^{2} +...+ t_{k-2}cc + t_{k-1}pp + t_{k} 
\end{equation}
Where $t_{1}, t_{2},...,t_{k}$ are the parameters of the surface which can be estimated by minimizing the least squared error. 
\begin{equation}
\underset{ t_{1},...,t_{k} }{\mathrm{min}} \sum_{i = 1}^{n} [f(p_i,cc_i,pp_i) - th_{i}]^2
\end{equation}

However, it under-fits the historical log severely. One of the good side of this modeling is that it provides a bitonic surface which is easy to explore. 

\textbf{\textit{(ii) Cubic regression:  }} A multivariate cubic surface can be represented as:
\begin{equation}
f(p,cc,pp) = t_{1}p^{3} + t_{2}cc^{3} + t_{3}pp^{3} + ...+ t_{k^\prime-1}pp + t_{k^\prime} 
\end{equation}
Corresponding least squared minimization problem can be defined as:
\begin{equation}
\underset{ t_{1},...,t_{k^\prime} }{\mathrm{min}} \sum_{i = 1}^{n} [f(p_i,cc_i,pp_i) - th_{i}]^2\\
\quad \mathrm{s.t.} \quad f(p,cc,pp) > 0
\end{equation}
This model also suffers from under-fitting the cluster data from historical logs. One way to resolve this under-fitting problem by introducing piecewise polynomials between the data samples with a guarantee of smoothness up to second derivatives.

\textbf{\textit{(iii) Piecewise cubic spline interpolation: }} We modeled throughput with cubic spline surface interpolation \cite{kincaid2009numerical}. Before introducing the interpolation method, we should explain the relationship among the parameters briefly. Concurrency and pipelining responsible for a total number of data streams during the transfer, whereas, pipelining is responsible for removing the delay imposed by small files. Due to their difference in characteristic, we model them separately. At first, we constructed a $2$-dimension cubic spline interpolation for $g(pp) = th$. Given a group of discrete points in $2$-dimension space $\{(pp_i,th_i)\}, i = 0,...,N$, the cubic spline interpolation is to construct the interpolant $g(pp) = th$ by using piecewise cubic polynomial $g_i(pp)$ to connect between the consecutive pair of points $(pp_i,th_i)$ and $(pp_{i+1},th_{i+1})$. The coefficients of cubic polynomials are constrained to guarantee the smoothness of the reconstructed curve. This is implemented by controlling the second derivatives since each piecewise relaxed cubic polynomial $g_i$ has zero second derivative at the endpoints. Now we can define each cubic polynomial piece as  

\begin{equation}
g_i(pp) = c_{i,0} + c_{i,1}pp + c_{i,2}pp^2 + c_{i,3}pp^3,\forall pp \in [pp_i,pp_{i+1}].
\end{equation}

Periodic boundaries can be assumed as $g(pp_{i+1}) = g(pp_i)$. Coefficients $c_{i,j}$, where $j = 1,2,3$, of piecewise polynomial $g_i(pp)$ contains $4(N-1)$ unknowns. We can have
\begin{equation}
g_i(pp_i) = th_i, \quad i = 1,...,N
\end{equation}
Hence, the $N$ continuity constraints of $g(pp)$ are as
\begin{equation} \label{eq:constraint2}
g_{i-1}(pp_i) = th_i = g_{i}(pp_i), \quad i = 2,...,N.
\end{equation}
We can get $(N-2)$ constraints from Equation (\ref{eq:constraint2}) as well. We can impose additional continuity constraints up to second derivatives. 
\begin{equation} \label{eq:constraint3}
\dfrac{d^2 g_{i-1}}{d^2 pp} (pp_i) = \dfrac{d^2 g_{i}}{d^2 pp} (pp_i), \quad i = 2,...,N
\end{equation}
We can get $2(N-2)$ constraints from Equation (\ref{eq:constraint3}). The boundary condition for relaxed spline could be written as,

\begin{equation} \label{eq:constraint4}
\dfrac{d^2 g}{d^2 pp} (pp_1) =  \dfrac{d^2 g}{d^2 pp} (pp_n) = 0
\end{equation}

So we have $N + (N-2) + 2(N-2) +2 = 4(N-1)$ constraints in hand. The coefficients can be computed by solving the system of linear equations. This example is extended to generate surface with two independent variables.

Then we modeled throughput as a piecewise cubic spline surface. It can be done by extending the above $2$-dimension cubic interpolation scheme to model the function of throughput. A short overview is given below:
We first fix the value of $pp$.
The throughput $f(p,pp,cc)$ then becomes $f_{pp}(p,cc)$
which is a surface in %$\Psi^2 \times \mathbb{R}$.
Data points can be represented as
$P = \{ (p_{i},cc_{j},th_{i,j}) | i = 1,2,\ldots,N, j = 1,2,\dots,M, (x_i,y_j) \in G \}$,
where $G$ is an $N \times M$ rectangle grid,
and each $(p_{i},cc_{j})$ denotes the grid point at $i$-th row and $j$-th column. The piecewise cubic interpolation method will construct
an interpolated cubic function $f_{r(i,j)}(p,cc)$
for each rectangle $r(i,j)$ from the grid $G$,
where $r(i,j)$ rectangle can be defined by $[p_i, p_{i+1}] \times [cc_j, cc_{j+1}]$.

$f_{r(i,j)}(p,cc)$ should fit the $th$-values of $P$ at the vertices of $r(i,j)$, \emph{i.e.}

\begin{gather*} 
f_{r(i,j)}(p_i,cc_j) = th_{i,j} \\
f_{r(i,j)}(p_{i+1},cc_j) = th_{i+1,j} \\
f_{r(i,j)}(p_i,cc_{j+1}) = th_{i,j+1} \\
f_{r(i,j)}(p_{i+1},cc_{j+1}) = th_{i+1,j+1} \\
\end{gather*} 

The interpolated cubic functions should also maintain smoothness at grid points. The continuity constraints are computed as an extension of   Equation (\ref{eq:constraint3}) and  (\ref{eq:constraint4}). 

\begin{figure}[t!]
    \centering
        \includegraphics[keepaspectratio=true,width=45mm]{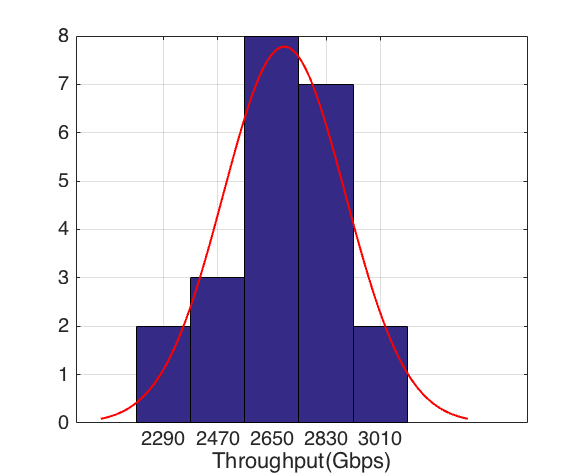}
        \includegraphics[keepaspectratio=true,width=44mm]{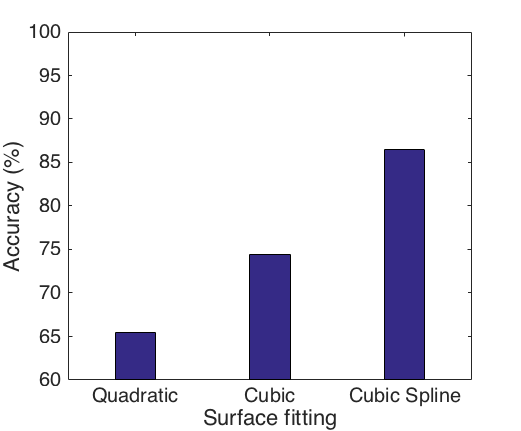}
    \caption{(a) Distribution of throughput values under similar external loads, b) Accuracy of different surface construction methods}
    \label{fig:surface_accuracy}
\end{figure}

After solving the  constrains (which is a linear system),
all the functions, $f_{rect(i,j)}(x,y)$, $i = 1,2,\ldots,N-1, j = 1,2,\dots,M-1$
form a piecewise smooth function which fixes data point set $P$.
Figure \ref{fig:3diinterpolation} shows the constructed piecewise cubic surfaces of throughput for different $cc$ and $p$ value. We can see from the graph surfaces for small files are more complex than the medium and large file. Figure \ref{fig:ppCurve} shows interpolated curve of throughput for different $pp$ values.
Figure \ref{fig:surface_accuracy}(b) shows the accuracy of different surface construction models. It can be seen that piecewise cubic spline outperforms all other models and achieves almost 85\% accuracy. 

Data transfer requests within the same cluster $C_{i}$ with same protocol parameter values might have a deviation from one another due to measurement errors and many other network uncertainties such as - different packet route in the network layer, minor queuing delay etc. 
%Figure \ref{fig:surface_accuracy}(a) shows that this deviation follows the Gaussian distribution. 
We define those data points with same protocol parameter entries as $\omega$. To model this deviation we have used a Gaussian confidence region around each constructed surface. The probability density function of a Gaussian distribution is: 

\begin{eqnarray}
&& p(\omega;\mu,\sigma) = \dfrac{1}{\sqrt{2\pi\sigma^2}}e^{-\dfrac{(\omega-\mu)^2}{2\sigma^2}}, \label{eq:confidence}\\
&& \mu = \frac{1}{N}\sum_{i=1}^N th_i, \label{eq:confidence:mean}\\
&& \delta = \sqrt{\frac{1}{N}\sum_{i=1}^N(th_i-\mu)^2}, \label{eq:confidence:var}
\end{eqnarray}
where $\mu$ is the mean and $\sigma$ is the standard deviation of the data distribution. Figure \ref{fig:surface_accuracy}(a) shows the data model for Gaussian distribution.

\subsubsection{Find maximal parameters}
\label{subsubsec:findargmax}
Very high protocol parameter values might overburden the system. For this reason, many systems set upper bound on those parameters. Therefore, the parameter search space has a bounded integer domain. Assuming $\beta$ is the upper bound of the parameters, cubic spline surface functions can be expressed as $f_i: \Psi^{3} \Rightarrow \mathbb{R}^{+}$, where $\Psi = \{1,2,..,\beta\}$. To find the surface maxima, we need to generate all local maxima of $F = \{f_1,...,f_p\}$. This is achieved by performing the second partial derivative test on each $f_k$ \cite{kincaid2009numerical}. The main idea is shown as below. 

First, we calculate the Hessian matrix of $f_k$ as

\begin{equation}
\label{eq:Hessian_fk}
H_k = J(\bigtriangledown f_k) =\left[ {\begin{array}{*{20}c}
   \dfrac{\partial^2 f_k}{\partial p^2} & \dfrac{\partial^2 f_k}{\partial p\ \partial cc} & \dfrac{\partial^2 f_k}{\partial p\ \partial pp}  \\
   ... & ... & ... \\
\dfrac{\partial^2 f_k}{\partial pp\ \partial p} & \dfrac{\partial^2 f_k}{\partial pp\ \partial cc} & \dfrac{\partial^2 f_k}{\partial pp^2}  \\  
 \end{array} } \right],
\end{equation}
where $J$ stands for the Jacobian matrix
\begin{equation}
J(F) = \left[ {\begin{array}{*{20}c}
   \dfrac{\partial f_1}{\partial p} & \dfrac{\partial f_1}{\partial cc} & \dfrac{\partial f_1}{\partial pp}  \\
   ... & ... & ... \\
\dfrac{\partial f_p}{\partial p} & \dfrac{\partial f_p}{\partial cc} & \dfrac{\partial f_p}{\partial pp}  \\  
 \end{array} } \right].
\end{equation}

Then we obtain the coordinates of all local maxima in $f_k$ by calculating the corresponding $\{p,pp,cc\}$'s such that $H_k(p,pp,cc)$ is negative definite. Hence, the set of local mxima of $f_k$ is obtained. Finally, the surface maxima is generated by taking the maximum among all local maxima sets of $F$. 

\subsubsection{Accounting for known contending transfers}
\label{subsubsec:knowncontendingtransfers}

Underlying TCP protocol tries to provide a fair share of bandwidth to all data streams concurrently transferring data. Assume we are analyzing a data transfer log entry $t_{p}$, any contending transfer in source or destination can have an impact on transfer request $t_{p}$. Known contending transfers are the ones present in the historical log. 

In a shared environment, there could be many transfers those are not explicitly logged, however, those unknown transfers could have an impact on the achievable throughput as well.  We can define the impact of those uncharted transfers as external load intensity, $I_{s}$, and model it with a simple heuristics:

\begin{equation}
I_s = \dfrac{bw - th_{out} }{bw}
\end{equation}

\subsubsection{Identify suitable sampling regions}
\label{subsubsec:unknownload}
Identifying the suitable sampling region is a crucial phase that helps online adaptive sampling module to converge faster. %From the previous phases, we know the cluster $C_{i,j,k}$ contains the throughput surfaces with different load intensities.
%Those surfaces interpolate throughput inside the bounded domains of the parameters. However, 
However, not all the regions on a surface are interesting. Many parameter coordinates of a surface are suboptimal. We are interested in regions which have better possibility of achieving high throughput. The regions containing distinguishable characteristics of the surfaces and containing the local maxima of those surfaces are more compelling. Exploring those regions could lead to a near-optimal solution much faster. Assume the cluster $C_{i}$ contains $\eta$ number of the surfaces that can be written as $S = {f_{1}, ... f_{\eta}}$. Now, we can extract the neighborhood with a predefined radius $r_d$ that contains maxima for all the surfaces in $S$. Assume the set $R_m$ contains all those neighborhood of maxima. We are also interested in regions where surfaces are clearly distinguishable. The goal is to find the regions where surfaces are maximally distant from one another. This problem can be formulated as a max-min problem. Selection can be done by taking the maximum of all pair shortest distance between the surfaces. To achieve that we perform uniform sampling $u = \{u_{1},...,u_{\gamma}\}$ from surface coordinate $(p,cc,pp)$  for surfaces in $S$. Therefore, $u$ could be written as 

\begin{equation}
u = \{u_{1},...,u_{\gamma}\} =  \{(p_i,cc_i,pp_i)\}^{\gamma}_{i=1} 
\end{equation}
We define $\Delta^{min}_{u_i}$ as the minimum distance between any two pair of surfaces that can be expressed as Equation (\ref{eq:region_selection}): 
\begin{equation}
\label{eq:region_selection}
\forall u_{k} \in u, \quad \Delta^{min}_{u_k} = \underset{ \forall i,j \in \{1,...,\eta\} }{\mathrm{min}} |f_{i}(u_k) - f_{j}(u_{k})| \quad \textrm{where} \quad i \neq j
\end{equation}
 
 After sorting the list in descending order we choose,$\lambda$ $(1<\lambda < k)$ number of initial samples from sorted list.  
 Assume the set of points we get after solving the equation (\ref{eq:region_selection}) is $R_c$. We define suitable sampling region as 

\begin{equation}
R_s = R_m \cup R_c
\end{equation}

During online analysis, we will use the region in $R_s$ to perform sample transfers.

\vspace{5mm}
\subsection{Adaptive Sampling Module}
\label{subsec:onlineadaptivesampling}
This module is initiated when a user starts a data transfer request. Adaptive sampling is dependent on online measurements of network characteristics. It is essential to assess the dynamic nature of the network that is helpful to find the optimal parameter settings. A sample transfer could be performed to see how much throughput it can achieve. However, a single sample transfer could be error prone and might not provide clear direction towards the optimal solution. Our algorithm adapts as it performs sample transfers by taking guidance from offline surface information. This approach can provide faster convergence. An overview of the module is presented in Algorithm (\ref{algo:onlinesampling}).  Online module queries the Offline analysis module with network information, such as - bandwidth, RTT, System information and the information about the dataset that user requested to transfer. Offline module finds the closest cluster and returns the throughput surfaces along with associated external load intensity information and suitable sampling region for each surface (Line 17,18). Then Online module sorts the surfaces in descending order based on external load intensity value.  
Adaptive sampling module takes the dataset that is needed to be transferred and starts performing sample transfers from the dataset. To perform first sample transfer, the algorithm chooses the surface with median load intensity, $f_{median}$. And perform the transfer with:

\begin{equation}
\theta_{s,median} = \{p,cc,pp\} = \mathrm{argmax}(f_{s,median})
\end{equation}

%%% Adaptive Sampling Algorithm:
%\IncMargin{1em}
\begin{algorithm}[t]
	\SetKw{in}{in}
	\SetKwData{Left}{left}\SetKwData{This}{this}
    \SetKwData{Up}{up}\SetKwData{Log}{Log}
    \SetKwData{ismedian}{$e_{s,median}$}\SetKwFunction{Median}{Median}
    \SetKwFunction{Union}{Union}
    \SetKwFunction{FindCompress}{FindCompress}
    \SetKwFunction{FindClosestSurface}{FindClosestSurface}
    \SetKwFunction{GetOptimalParam}{GetOptimalParam}
    \SetKwFunction{DataTransfer}{DataTransfer}
    \SetKwFunction{AdaptiveSampling}{AdaptiveSampling}
    \SetKwFunction{append}{append}
    \SetKwFunction{remove}{remove}
    \SetKwFunction{QueryDB}{QueryDB}
    \SetKwFunction{Sort}{Sort}
    \SetKwFunction{GetSamples}{GetSamples}
	\SetKwInOut{Input}{input}\SetKwInOut{Output}{output}
    \tcp{Source, $EP_{s}$, Destination, $EP_{d}$, Round trip time, $rtt$, Bandwidth, $bw$}
	\Input{Data arguments, $data\_args = \{Dataset , avg\_file\_size, num\_files\}$,
           network arguments, $net\_args = \{EP_s, EP_d, rtt, bw\}$,  
           transfer node arguments, $node\_args = \{num\_nodes,cores, memory, NIC\_speed \}$}
	\Output{Optimal transfer rate, $th_{opt}$}
	\BlankLine
	%\emph{special treatment of the first line}\;
  \SetKwProg{proc}{procedure}{}{}
   \proc{\AdaptiveSampling{$F_s$, $R_s$, $I_s$} }{
    	$D_{s}$ $\leftarrow$ \GetSamples{$Dataset$}  \\
        
    	\ismedian  $\leftarrow$ \Median{$I_s$} \\
        $f_{s,median}$ $\leftarrow$ $F_{s}[e_{s,median}]$ \\
        $\theta_{s,median}$, $th_{hist}$ $\leftarrow$ \GetOptimalParam{$f_{s,median}$} \\
        $th_{cur}$ $\leftarrow$ \DataTransfer{$D_{s,1}$,$p_{s,median}$} \\
        \Log.\append{$net\_args$, $D_{s,i}$,$p_{s,median}$,$th_{cur}$} \\
        $D_{s}$.\remove{$D_{s,1}$} \\
        \For{$D_{s,i}$ \in $D_s$ } {
              \If{$th_{cur}$ $\neq$ $th_{hist}.confidence\_bound$} { 
                   $f_{s,cur}$ $\leftarrow$ \FindClosestSurface{$th_cur$} \\
                   $p_{s,cur}$, $th_{hist}$ $\leftarrow$ \GetOptimalParam{$f_{s,curr}$} \\
                   $th_{cur}$ $\leftarrow$ \DataTransfer{$D_{s,i}$,$p_{s,cur}$} \\
                   \Log.\append{$net\_args$, $D_{s,i}$,$p_{s,cur}$,$th_{cur}$}
              }
        }
    }
    
    $F_s$, $R_s$, $I_s$ $\leftarrow$ \QueryDB{$data\_args$,$net\_args$} \\
	$F{'}_{s}$ $\leftarrow$ \Sort{$F_s$,$I_s$} \\
    \tcp{Set of surfaces, $F_{s}$, Sampling region, $R_{s,k}$, Load intensity, $I_s$ }
	\AdaptiveSampling{$F^{'}_{s}$, $R_{s,k}$, $I_s$}
\caption{Online Sampling}
\label{algo:onlinesampling}
\end{algorithm}
%\DecMargin{1em}

which is already precomputed during offline analysis and can be found in sampling region. Achieved throughput value for the transfer is recorded (Line 2-6). If the achieved throughput is inside the surface confidence bound at point $\theta_s,median$, then the algorithm continues to transfer rest of the data set chunk by chunk. However, if the achieved throughput is outside the confidence bound, that means the current surface is not representing the external load of the network. If achieved throughput is higher than surface maxima, that means current network load is lighter than the load associated with the surface. Therefore, the algorithm searches the surfaces with lower load intensity tags and find the closest one and perform second sample transfer with parameters of newly found surface maxima. In this way, the algorithm can get rid of half the surfaces at each transfer. At the point of convergence, our algorithm takes the rest of the dataset and starts the transfer process. Changing parameters in real time is expensive. For example, if a $cc$ value changes from 2 to 4, this algorithm has to open two more server processes, initialize resources. These new processes have to go through TCP Slow start phase as well. Therefore, the algorithm tries to minimize the initial sampling transfers by the adaptive approach. For very large scale transfers, when data transfer happens for a long period of time, external traffic could change during the transfer. If algorithm detects such deviation, it uses most recently achieved throughput value to choose the suitable surface and changes the transfer parameters.   

\iffalse
\begin{figure}[t]
	\begin{centering}
			\includegraphics[keepaspectratio=true,angle=0,width=80mm]{figures/Dst.pdf}
	\end{centering}
    \vspace{-3mm}
\caption{Different types of known contending transfers.} \label{fig:srcDst}
\end{figure}
\fi

\section{Experiments}
\label{sec:experiments}
%% Description of Dataset:
In our study, we have used real production level Globus data transfer logs \cite{globusonline} to analyze network and throughput behaviors, end system characteristics, peak and off-peak hour transfer loads. For our experiments, we used XSEDE, a collection of high-performance computing resources connected with high-speed WAN and our DIDCLAB testbed. On XSEDE, we performed data transfers between Stampede at Texas Advanced Computing Center (TACC) and Gordon cluster at San Diego Supercomputing Center (SDSC). Table \ref{table:specfications} shows the system and network specifications of our experimental environment.  

%% Table: specifications:
\begin{table}[]
\centering
\caption{System specification of our experimental environment}
\label{table:specfications}
\begin{tabular}{|l|l|l|c|c|}
\hline
                                                           & \multicolumn{2}{c|}{XSEDE}                              & \multicolumn{2}{c|}{DIDCLAB}                               \\ \hline
                                                           & Stampede                   & Gordon                     & \multicolumn{1}{l|}{WS-10} & \multicolumn{1}{l|}{Evenstar} \\ \hline
\hline
Cores                                                      &                            &                            & 8                          & 4                             \\ \hline
Memory                                                     &                            &                            & 10 GB                      & 4 GB                          \\ \hline
Bandwidth                                                  & \multicolumn{2}{c|}{10 Gbps}                            & \multicolumn{2}{c|}{1 Gbps}                                \\ \hline
RTT                                                        & \multicolumn{2}{c|}{40 ms}                              & \multicolumn{2}{c|}{0.2 ms}                                \\ \hline
\begin{tabular}[c]{@{}l@{}}TCP \\ Buffer size\end{tabular} & \multicolumn{1}{c|}{48 MB} & \multicolumn{1}{c|}{48 MB} & 10 MB                      & 10 MB                         \\ \hline
\begin{tabular}[c]{@{}l@{}}Disk \\ Bandwidth\end{tabular}  & 1200 MB                    & 1200 MB                    & 90 MB                      & 90 MB                         \\ \hline
\end{tabular}
\end{table}

%% Different models:
We compared our results with many other existing models, such as - (1) Static models - Globus (GO) \cite{allen:2012}, Static Parameters (SP) \cite{Nine:2015ANN},  (2) Heuristic Model - Single Chunk (SC) \cite{engin:dynamicTuning}, (3) Dynamic model -HARP \cite{Engin2016}, ANN+OT \cite{Nine:2015ANN}, (4) Mathematical model - Nelder-Mead Tuner (NMT) \cite{Ian-bala2016}. Globus uses different static parameter settings for different types of file sizes. SC  also makes parameter decision based on dataset characteristics and network matrices. It asks the user to provide an upper limit for concurrency value. SC does not exceed that limit. HARP uses heuristics to perform a sample transfer. Then the model performs online optimization to get suitable parameters and starts transferring the rest of the dataset. Online optimization is expensive and wasteful as it needs to be performed each time, even for similar transfer requests. ANN+OT  learns the throughput for each transfer requests from the historical log. When a new transfer request comes, model asks the machine learning module for suitable parameters to perform first sample transfer. Then it uses recent transfer history to model current load and tune parameter accordingly. The model only relies on historical data and always tends to choose the maxima from historical log rather than the global one.  Nelder-Mead Tuner implements a direct search optimization which does not consider any historical analysis, rather tries to reach optimal point using reflection and expansion operation. We tested those models three real production level networks - (1) Between XSEDE nodes, (2) DIDCLab testbed, and (3) Between DIDCLab and XSEDE nodes.

\begin{figure*}[t] 
\centering
		\includegraphics[scale=0.3]{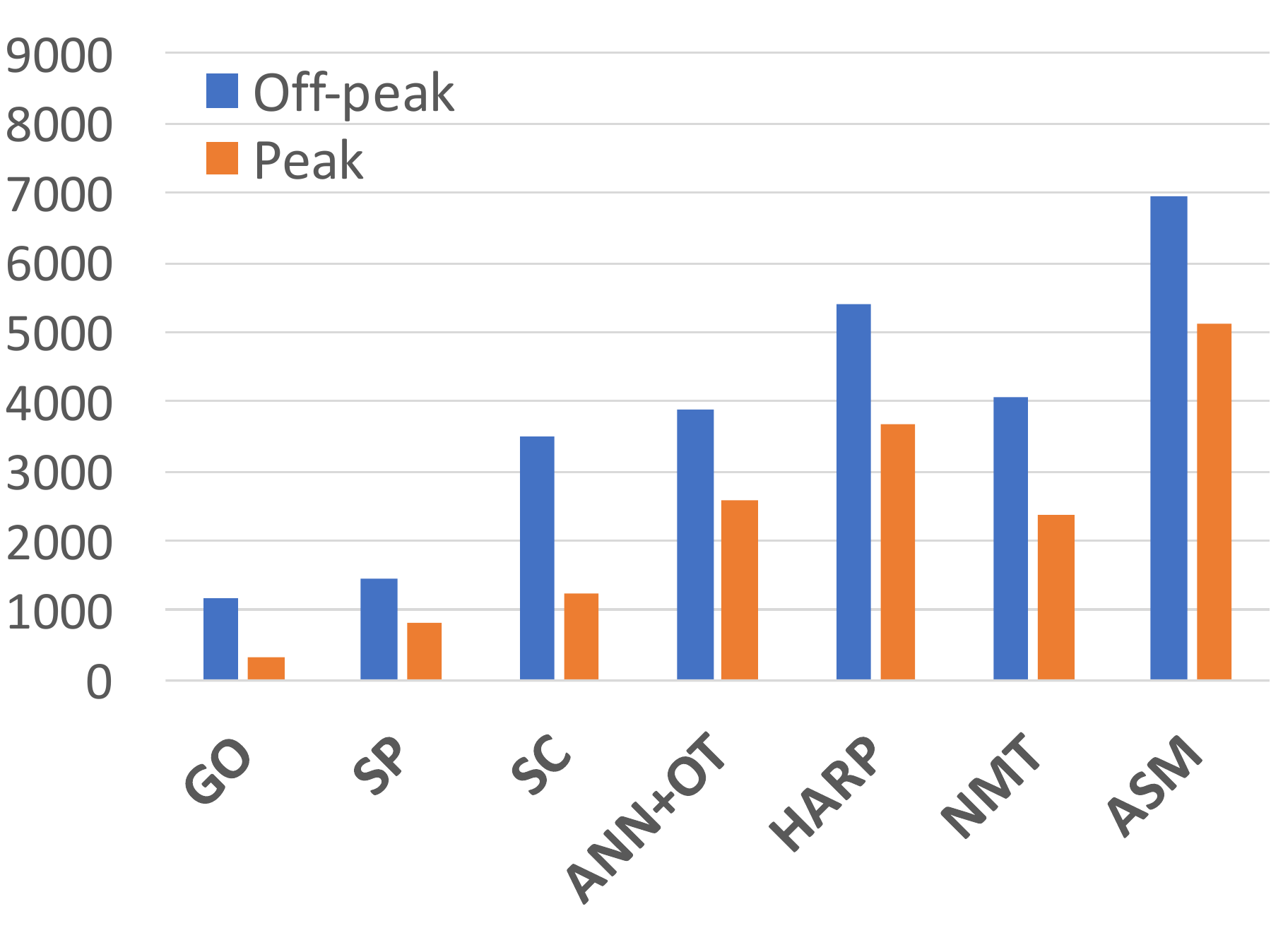}
		\includegraphics[scale=0.3]{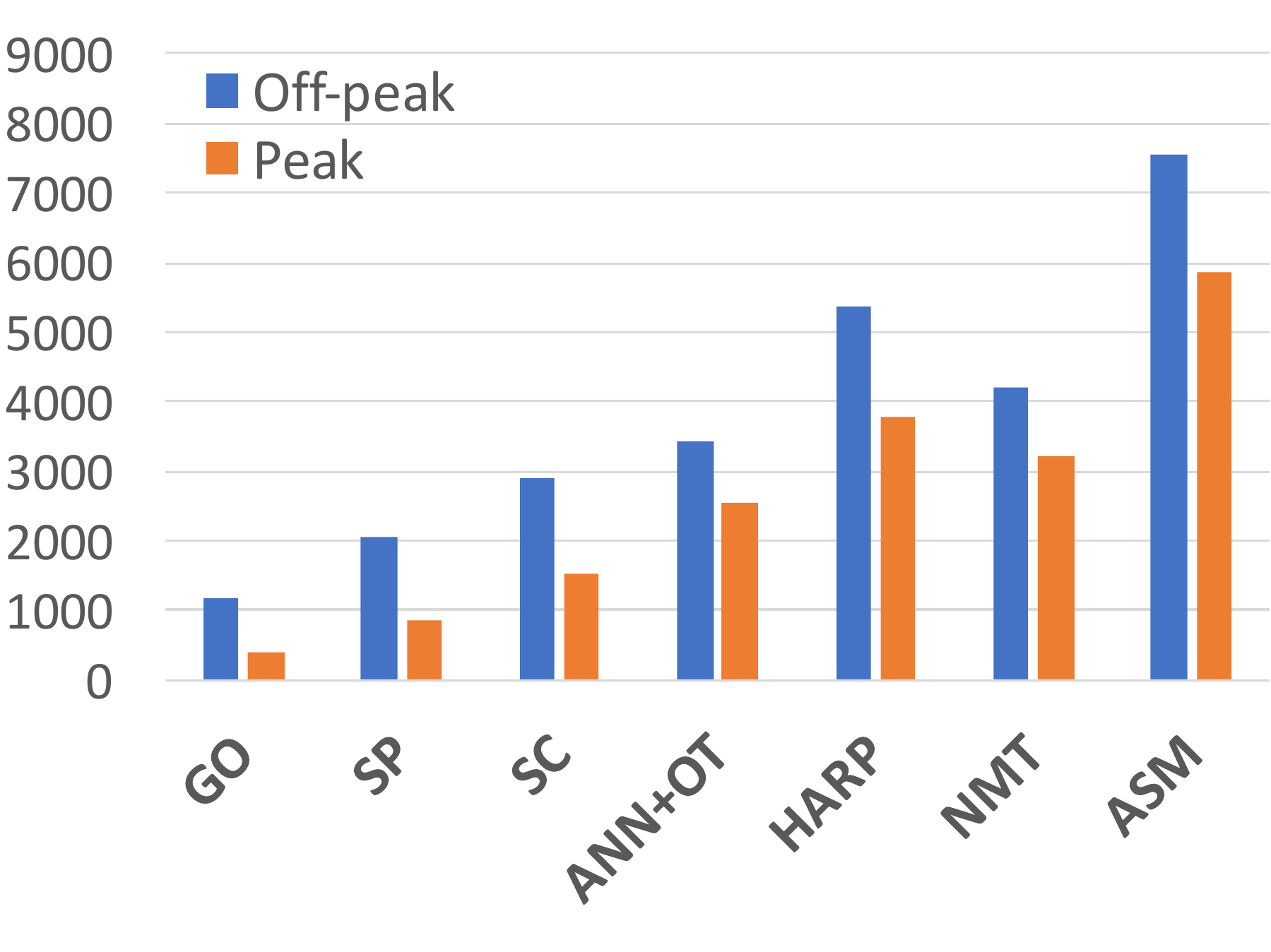}
		\includegraphics[scale=0.3]{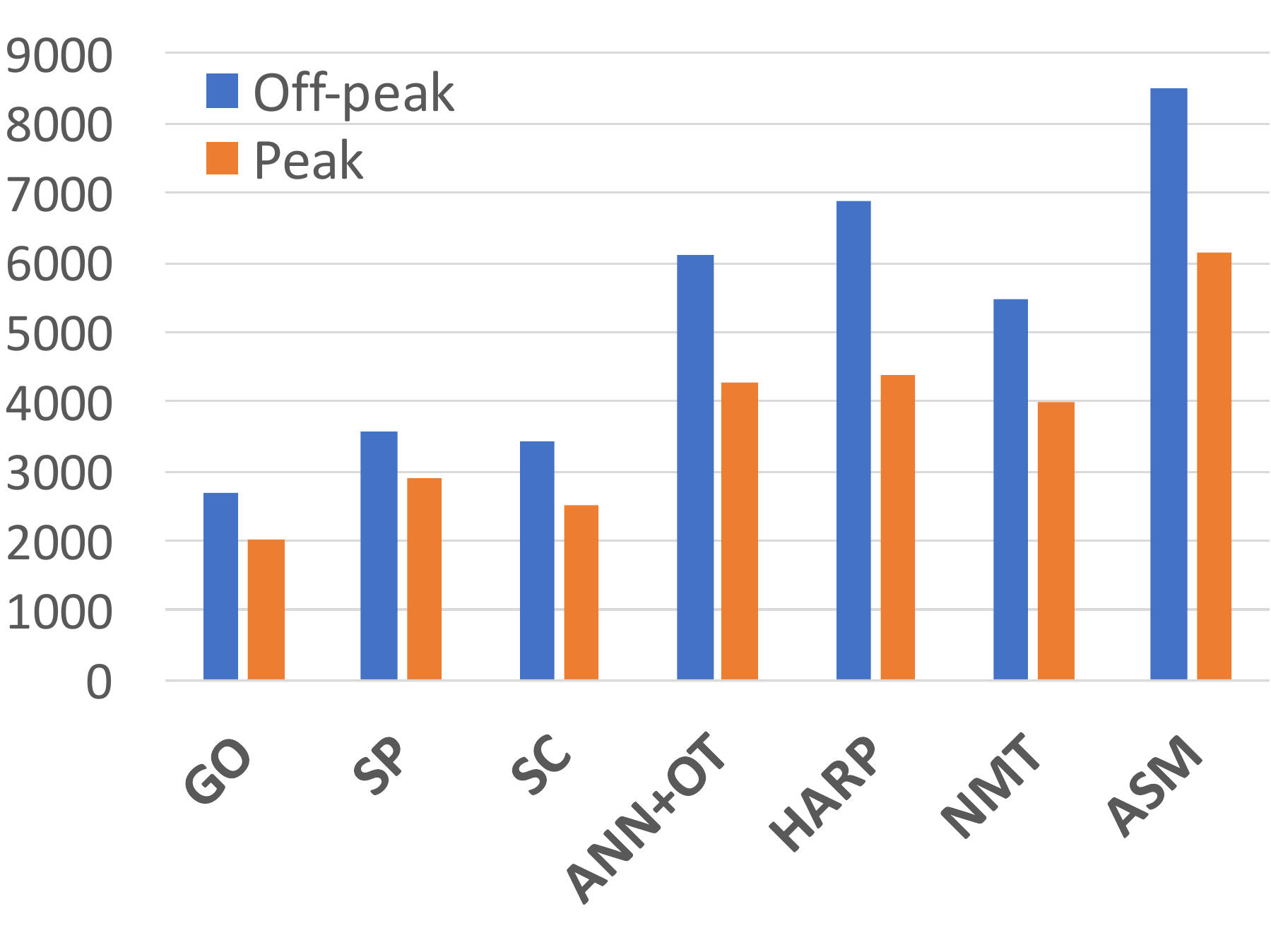}
		\includegraphics[scale=0.3]{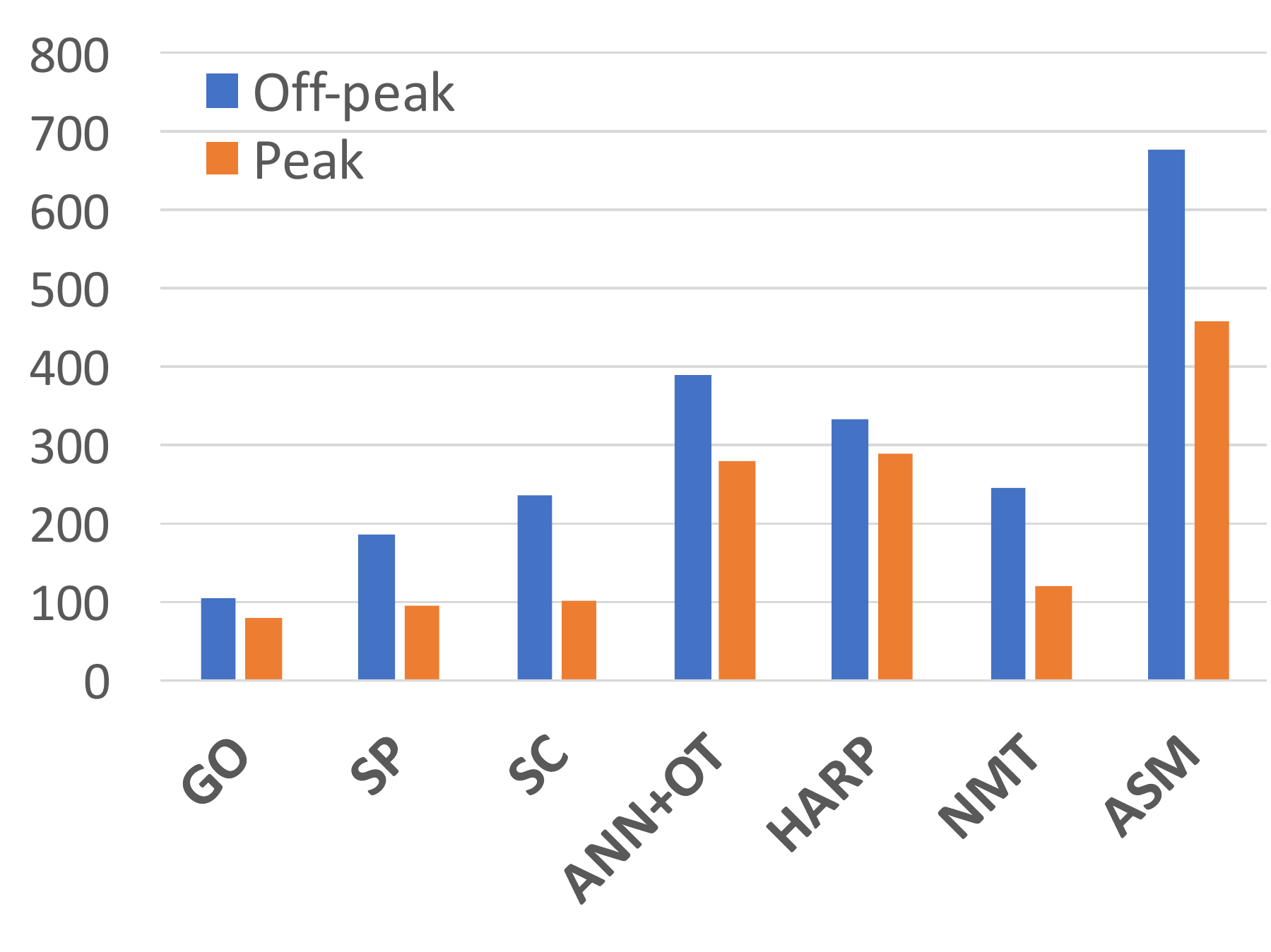}
		\includegraphics[scale=0.3]{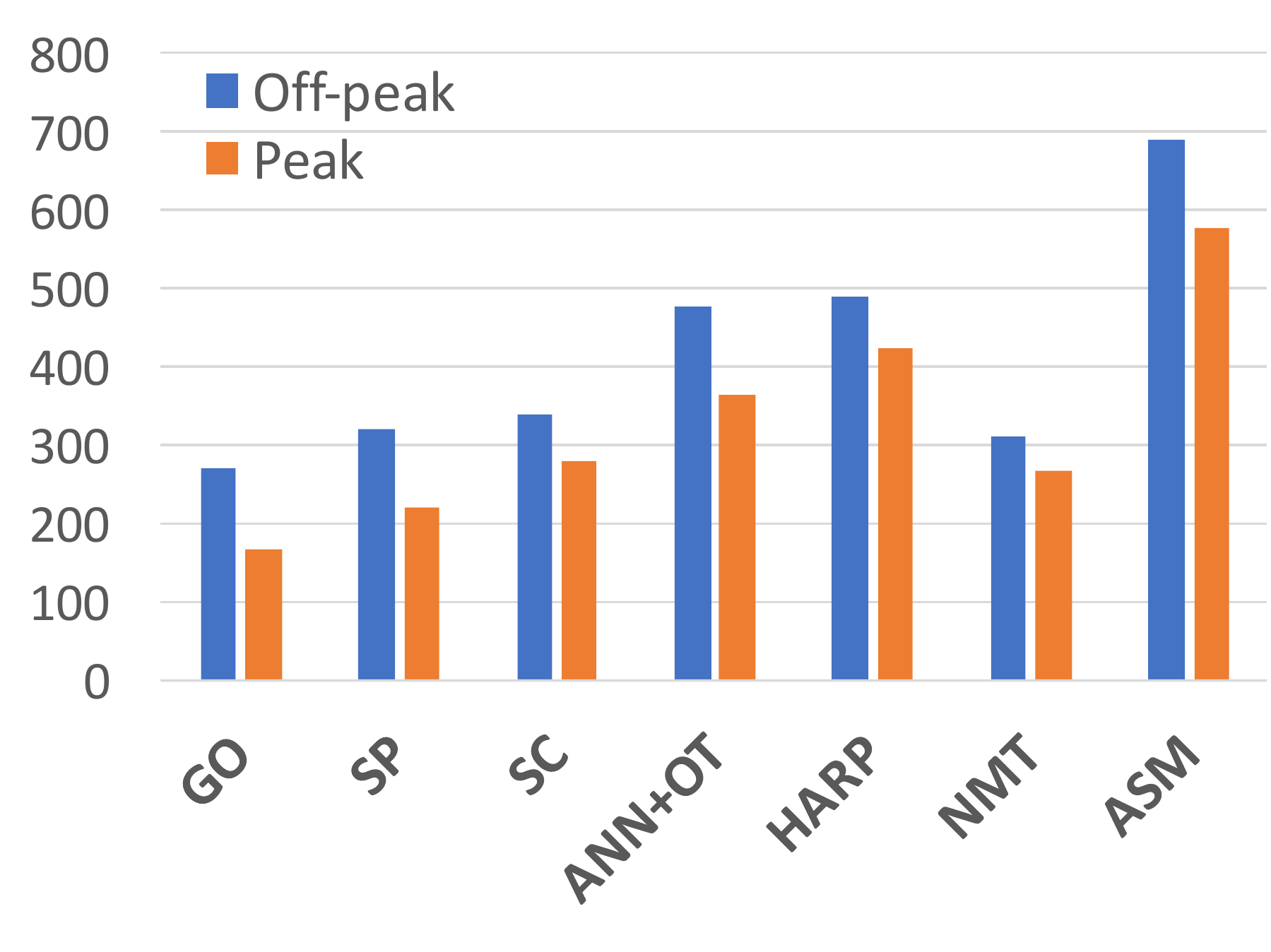}
		\includegraphics[scale=0.3]{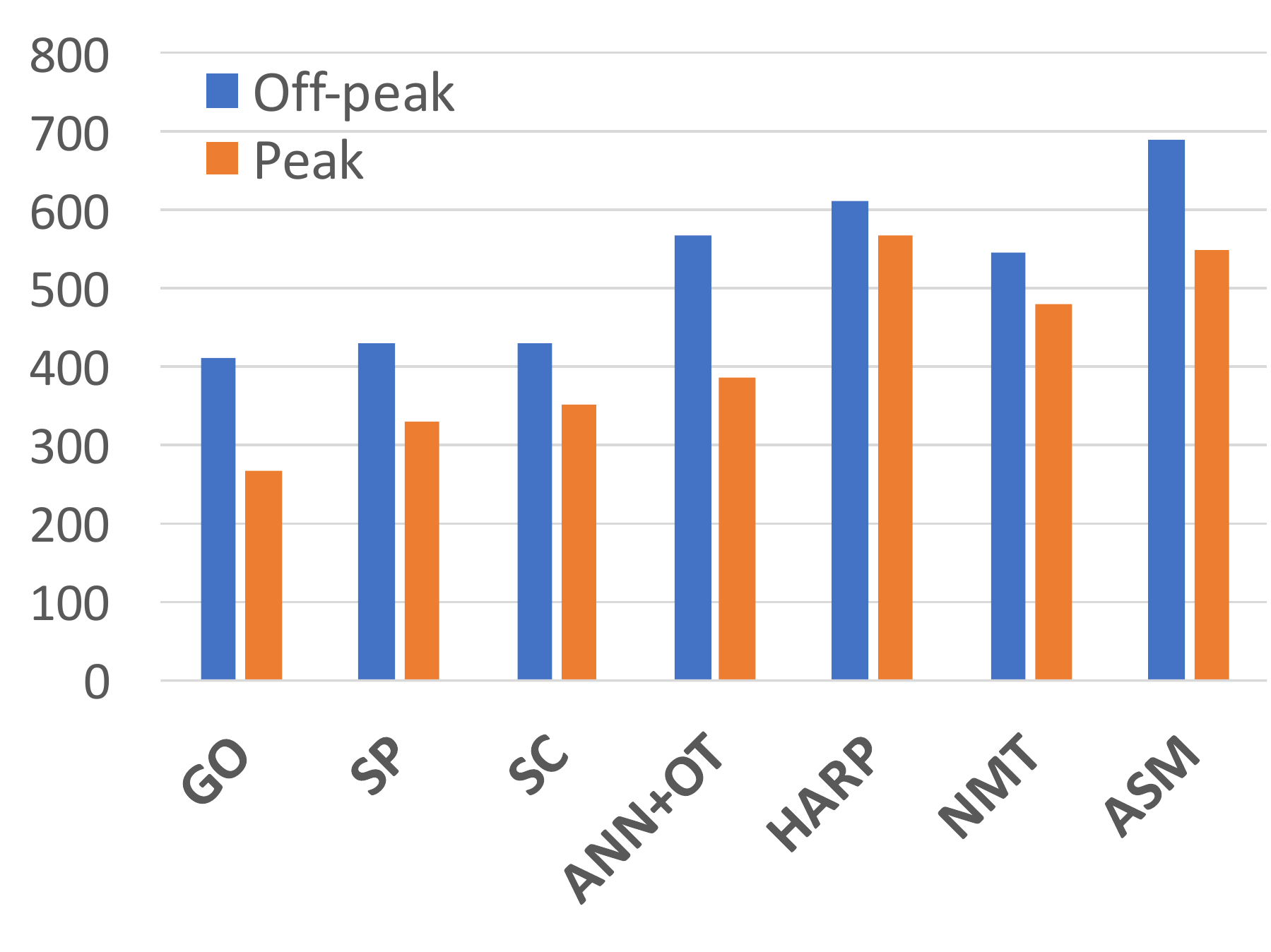}
		\includegraphics[scale=0.3]{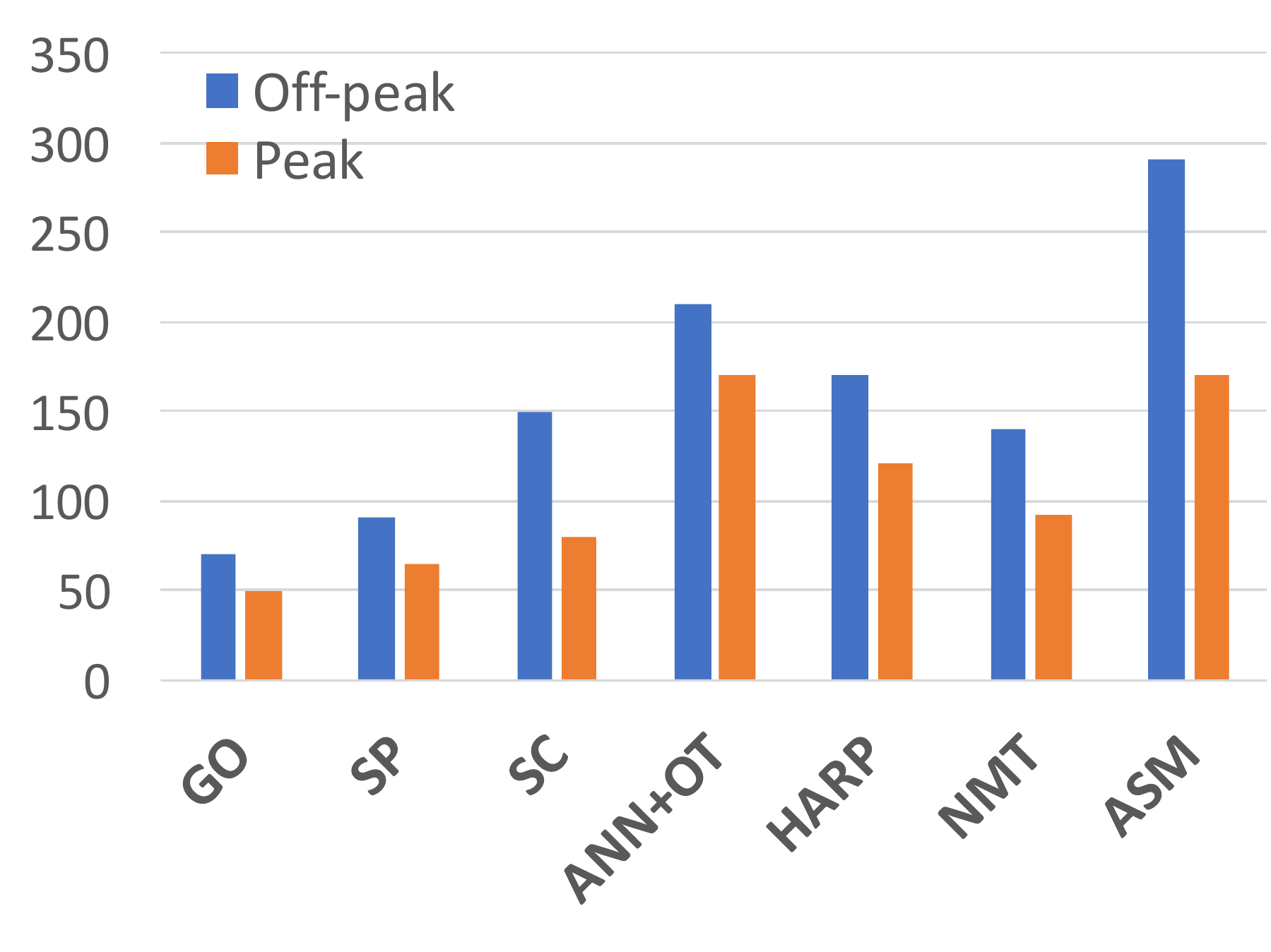}
		\includegraphics[scale=0.3]{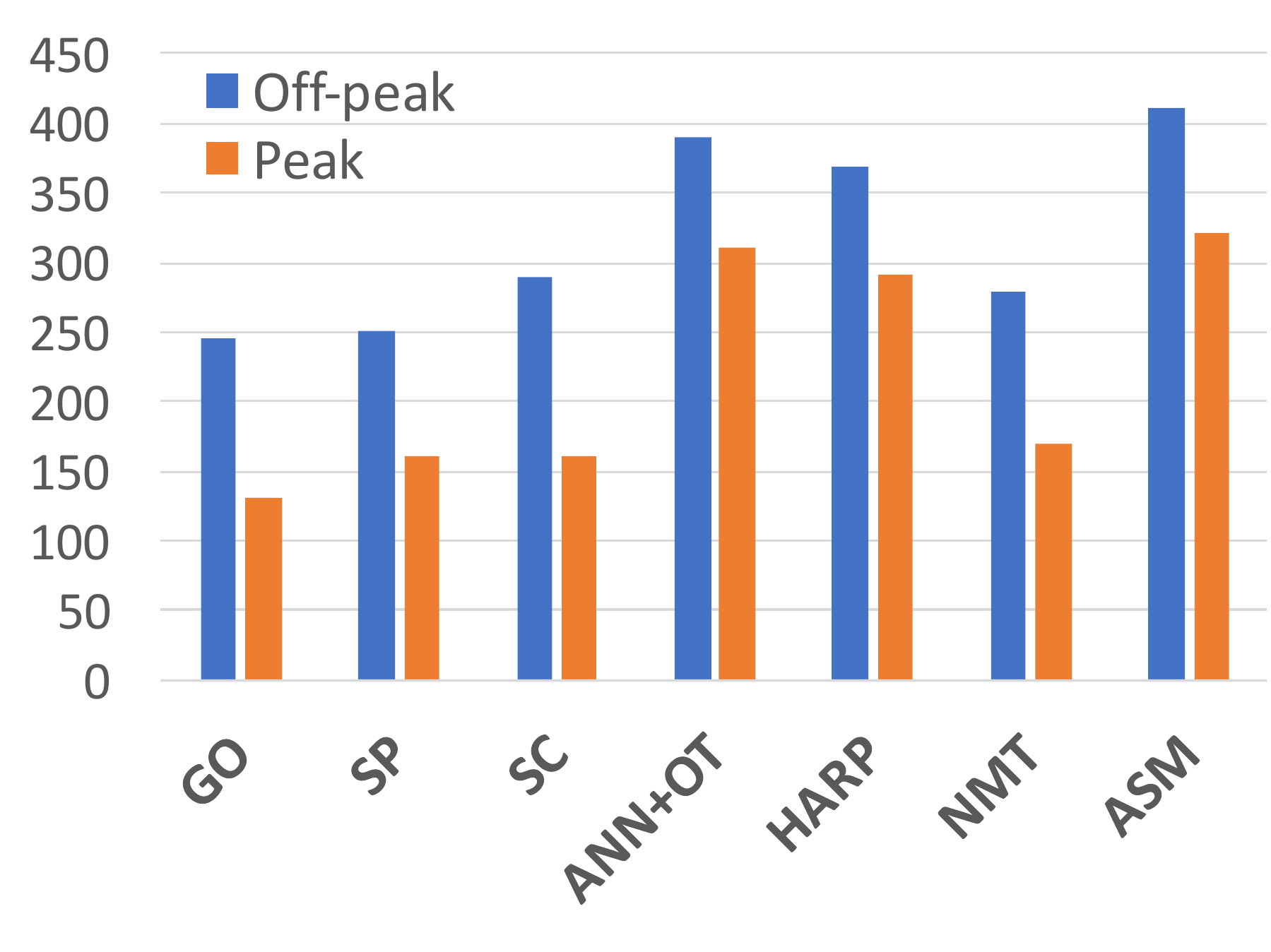}
		\includegraphics[scale=0.3]{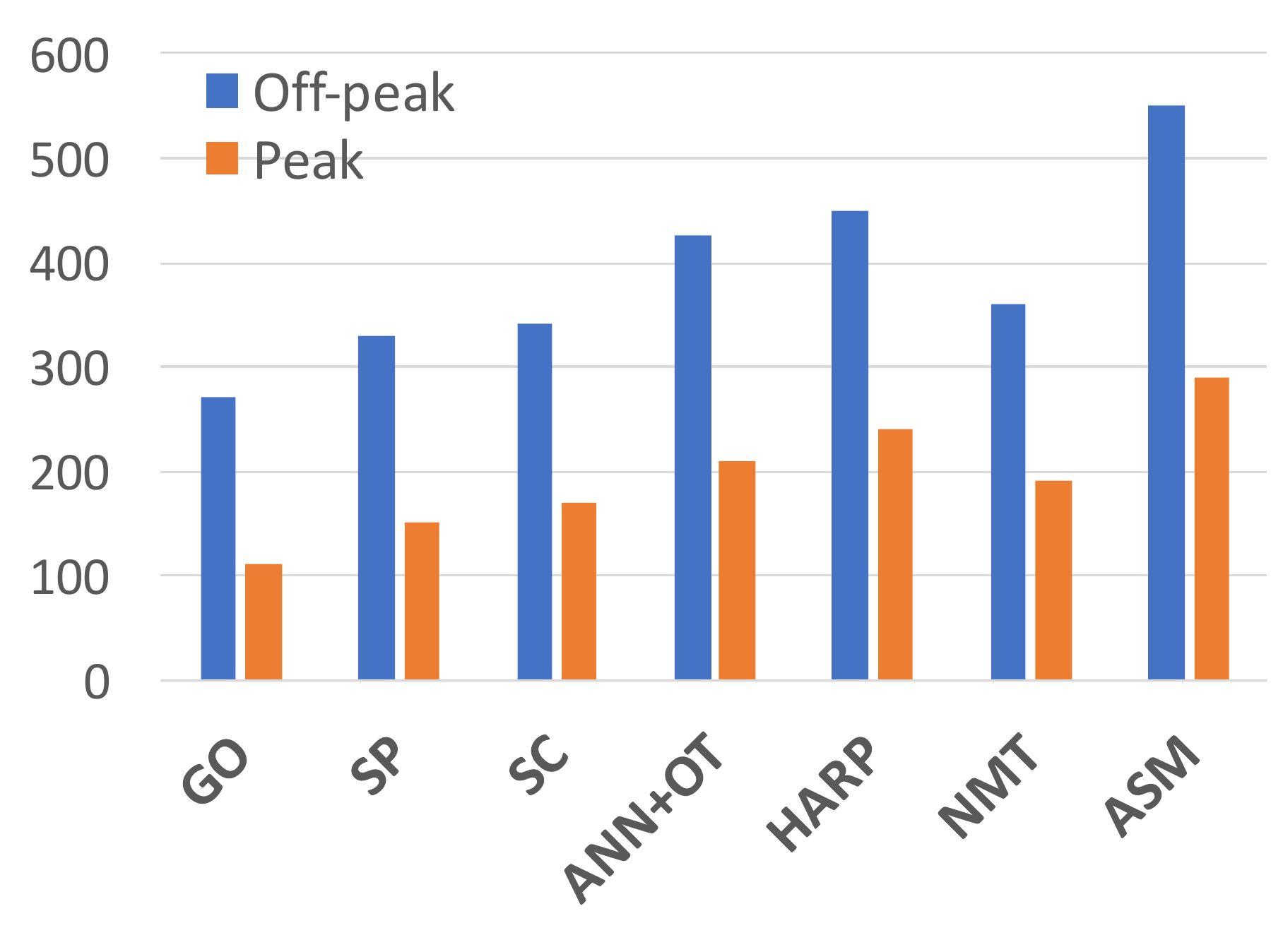}
\caption{Achievable throughput (Gbps) in our experiments performed in various environments for different file sizes.  }
\label{fig:throughputresults}
\end{figure*}

\subsection{Transfers between XSEDE nodes}
\label{subsec:transfers_between_xsede_nodes}
We tested our model with data transfer requests those are completely different from the historical logs used in the model. To ensure that we computed the list of all unique transfers and split the list as 70\% for training the model and 30\% for test purpose. We also evaluated our model on both peak and off-peak hours to measure performance under different external load conditions. Achievable throughput is also highly dependent on the average file sizes of the dataset. For example, a dataset with average file size 2 MB and 4 MB achieve two different level of throughputs than dataset with average file size 100MB or 200MB. %Unlike SC, our model does not make any hard decision on average file size. Offline clustering make those groups in a completely unsupervised manner based on historical data.% 
To see how accurate our model is for different types of average file sizes, we partitioned test data transfer requests into three groups - small, medium and large. Then we compared average achievable throughput so that we can evaluate the model in a more fine-grained way. Figure \ref{fig:throughputresults} (a-c) shows the comparison among above mentioned models and our proposed models. For XSEDE transfers, Globus can achieve up to 2700 Mbps throughput for large file transfers during off-peak hours, however, achieved throughputs are significantly lower for the medium and small dataset. Static parameters from \cite{Nine:2015ANN} can achieve almost 100\% increase in throughput compared to Globus for medium files.It also outperforms GO during small and large dataset transfers. It uses the knowledge from historical data to set those parameters. For example, consider two identical transfers are performed over similar external loads. First transfer uses $cc = 4$ and $p = 4$ and the second one uses $cc = 8$ and $p = 2$. Even though in both cases, the total number of TCP streams is 16, the later one achieves significantly higher throughput than the first one. This is because in the later case, eight GridFTP server processes are pushing data through the network instead of four. For small files, higher $cc$ and $pp$ value up to a certain extent work well. However, in a dynamic environment, such strict static settings might lead to under-utilization or very high value could lead to severe packet loss and queuing delay. SC provides around 140\% throughput increase for small files compare to Static parameters. The user-provided upper limit for concurrency is set to 10. We can see it performs better than static parameters for all three types of dataset groups. SC uses information like average file sizes, the number of files, network round trip time, TCP Buffer size, network bandwidth etc.  All these information help SC to adapt to the network environment.
However, for large files, the performance increase is negligible. This is because SC sets concurrency to user defined value 10 which is close to Static parameter concurrency level 8.
ANN+OT outperforms SC in every single file category during off-peak and peak hours. However, for small files, the throughput increase is only 11\% during the off-peak hour. This is the effect of similar parameters choice of two models. 
During peak hour, ANN provides almost 38\% throughput increase for small files, 65\% increase for medium files and 70\% increase.
That is because, ANN+OT performs online transfers to model current network load, where SC is agnostic towards network traffic. 
However, HARP outperforms ANN+OT and NMT for all three types of datasets . HARP's better performance relies on expensive online optimization. ANN+OT excessively dependent on sample maxima. However, for large datasets, the performance of HARP during peak hours is pretty similar to ANN+OT. 
Even though NMT  ultimately reaches to global maxima, it suffers during peak period due to its slow convergence. Each time this algorithm changes the parameters, it has to stop the globus-url-copy command and has to start the command with new parameters, which comes with expensive initialization and TCP slow start phase. Sub-optimal parameter settings during convergence period hurt its overall performance, yet it is free from expensive online optimization. We can observe its degraded performance during the peak hour period. However, it outperforms all the static models along with SC. It actually proves the power of online sampling over the heuristic models. 
Adaptive Sampling Module (ASM) outperforms all other models, even the nearest one (HARP) by almost 40\% for medium datasets. For the large dataset, our model outperforms HARP by 23\%. In case of the small dataset, the performance increase is 29\%.  Adaptive sampling solves the slow convergence problem with the more accurate pre-constructed representation of throughput surfaces. Our model also gets rid off all the surface regions those proved suboptimal for different background traffic. Moreover, it has a fast online module with adaptive sampling that can converge faster and reduces the suboptimal convergence time. Moreover, our model obtains more impressive performance during peak hours. It outperforms HARP by almost 38\%, 55\%, and 39\% for small, medium, and large datasets respectively. Peak hour periods are challenging to model, and the result shows that our offline analysis is resilient enough to achieve better results in such network environment, with the help of adaptive sampling module.

\begin{figure}[t]
\begin{centering}
\includegraphics[keepaspectratio=true,angle=0,width=80mm]{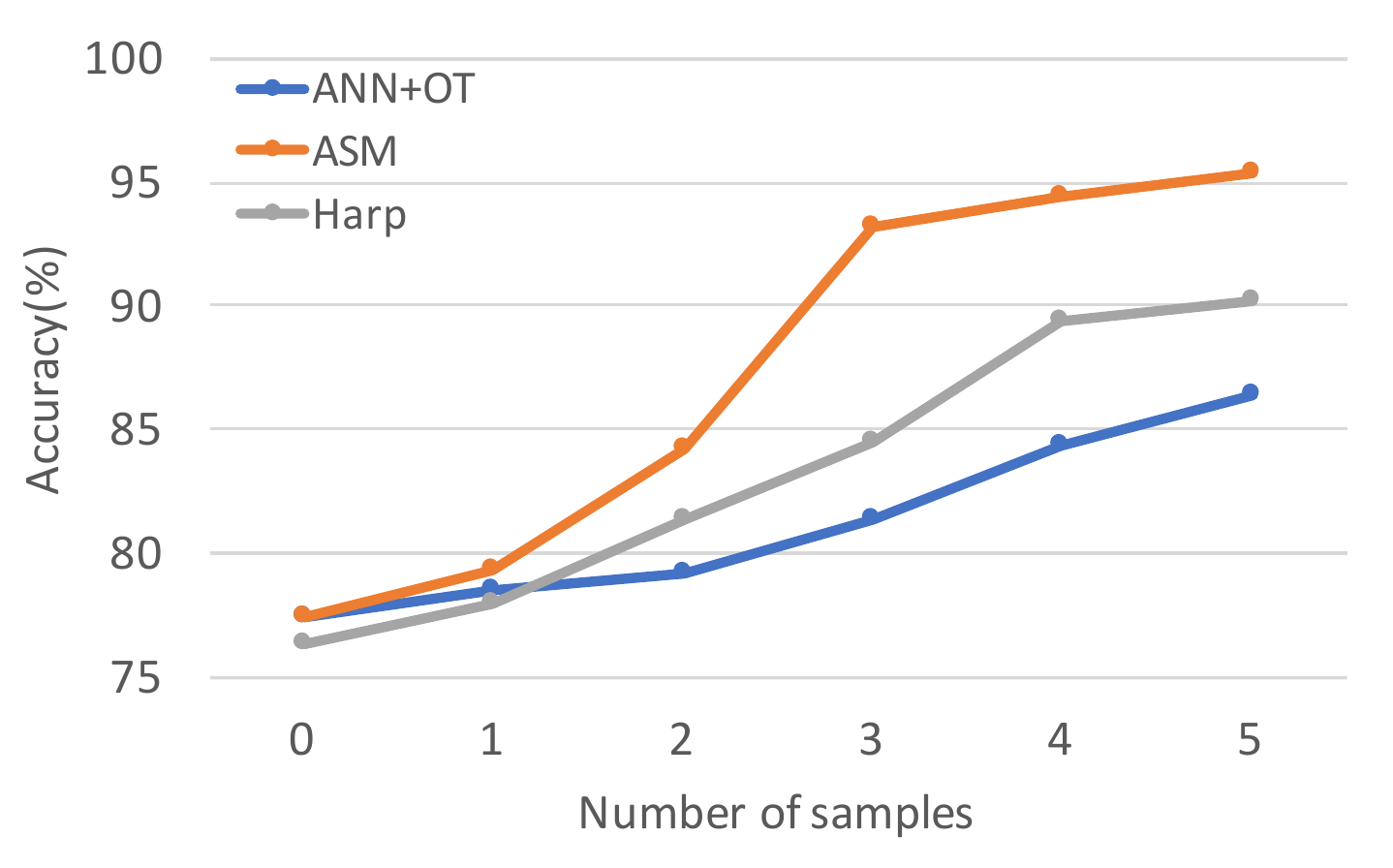}
	\end{centering}
    \vspace{-3mm}
\caption{Prediction accuracy of different models with respect to number of sample transfers (those uses online sampling). } \label{fig:predictionaccuracy}
\end{figure}

\subsection{DIDCLab testbed}
\label{subsec:didclab_testbed}
Figure \ref{fig:throughputresults}(d-f) shows the throughput performance of different models in our DIDCLAB testbed. As it is connected with University LAN, peak hour is experienced from 11 am to 3 pm. And night time is mostly the off-peak hour. Globus online better results on the large dataset, however, performs poorly during small and medium transfers. Static Parameters (SP) outperforms GO by almost 100\% during small file transfers. The performance difference is negligible during large file transfers. In this experimental setting, achievable throughput is actually bounded by disk speed. As single chunk is unaware of disk bottleneck, its parameters become suboptimal and exhibit similar results than SP during peak hours. Because of the historical analysis, ANN+OT can outperform all the static and heuristic models. For small files during peak hour ANN+OT can achieve $3\times$ throughput improvement. On the other hand, during the small dataset transfers, HARP shows performance degradation compare to ANN+OT. After the investigation, we figured out that some cases sample transfer finished during the TCP slow start phase with low throughput, which could mislead the online optimizer and eventually lead to performance degradation. However, HARP outperforms slightly ANN+OT during medium and large dataset transfers. In this environment, direct search shows interesting results. As it is unaware of disk bottleneck, convergence time takes longer than expected. The performance of medium and small dataset hurts the most, as a big portion of the data gets transferred during convergence period. That is why it shows similar results to static models during both peak and off-peak hours. Our model (ASM) outperforms all the existing models. It achieves 100\% performance improvement over HARP during small file transfers during off-peak hours. It outperforms HARP by 41\% during medium dataset transfers. However, for large files, the performance improvement is only 13\% and during peak hours HARP actually does slightly better than our model. HARP's performance basically depends on its regression accuracy, in this case, we observed high regression accuracy which leads to better performance. However, the rest of the experiments suggest that in this case, HARP gets "lucky".
\begin{figure}[t]
\begin{centering}
\includegraphics[keepaspectratio=true,angle=0,width=80mm]{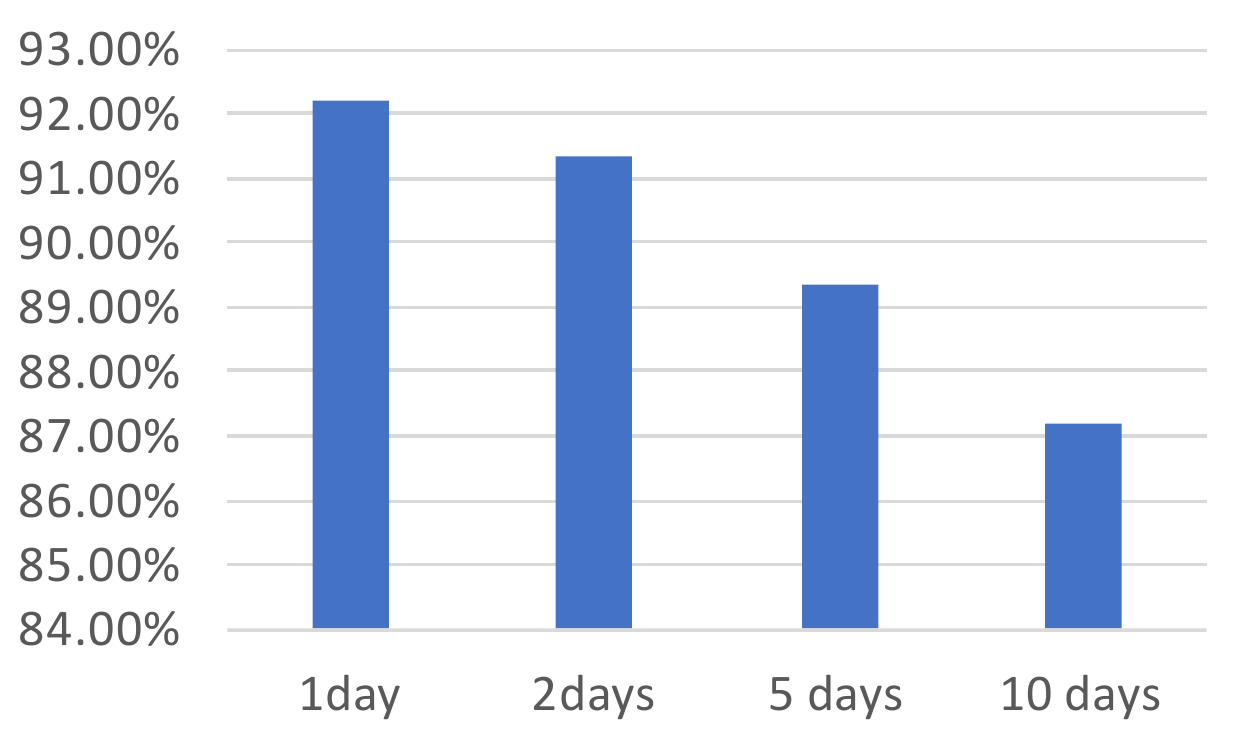}
	\end{centering}
    \vspace{-3mm}
\caption{ Model accuracy over periodic offline analysis.} \label{fig:periodicaccuracy}
\end{figure}

\subsection{Between DIDCLab and XSEDE}
\label{subsec:between_didclab_and_xsede}

In Figure \ref{fig:throughputresults} (g-i), we reported the transfer performance among the models on DIDCLAB to Gordon network. This is over the Internet connection which makes it more challenging. During off-peak period, GO performs better during medium and large data transfers rather than small transfers. Static parameters do better than GO during all the cases, except offline medium dataset transfers. Single chunk outperforms Static Parameters (SP) by 50\% during off-peak hours. As expected ANN+OT does better than heuristic and static models. Surprisingly, it performs better than HARP during small and medium file transfers. For medium file transfers its performance is very close to our model. We observed high neural network accuracy for predicting parameters that lead to better performance. This network environment is proved challenging for NMT, as unpredictable peak hour leads to slow convergence. Our model performed better than all the mentioned models. For small dataset, our model outperforms its closest competitor ANN+OT by 38\%. It outperforms HARP by 22\% during large dataset transfers.

Our online module needs almost constant time to agree on the parameters. Among the existing models that we have tested so far, only HARP uses the online optimization which could be expensive, however, rest of the models can perform transfers in constant time. 

Among the above-mentioned models, static, heuristics, and mathematical optimization models do not require any historical analysis, however, our model requires extra historical analysis. Therefore, a natural question would be, how often do we have to perform the offline analysis? The answer is, we do not need to perform offline analysis before every single data transfer requests, rather it can be done periodically. Figure \ref{fig:periodicaccuracy} shows the impact of offline analysis frequency on the accuracy of the model. Offline analysis performed once a day is enough to reach 92\% accuracy. Model accuracy decreases slightly to 87\% even for cases, where offline analysis is performed once in 10 days. This shows the model could converge faster, even when offline analysis are performed 10 days apart.

Adaptive Sampling Module(ASM) performs online sampling and uses the network information to query the offline analysis for optimal parameters along with achievable throughput, $T_{predict}$. the optimal parameters are used for next sample transfer. Then we measure the actually achieved throughput, $T_{achieved}$. As our model converge $T_achieved$ gradually gets closer to the $T_predict$. To measure the accuracy of the model we used the following metric:

\begin{equation}
\mathrm{Accuracy} = \dfrac{|T_{achieved }- T_{predict}|}{T_{predict}} \times 100
\end{equation}

Figure \ref{fig:predictionaccuracy} shows a comparison of the accuracy of throughput prediction models. HARP can reach up to 85\% with 3 sample transfers along with high online computation overhead. ANN+OT can reach 87.32\% accuracy. Our model achieves almost 93\% accuracy with three sample transfers for any types of Dataset and then it saturates. It shows that our offline cubic spline interpolation can model the network more accurately and adaptive sampling can ensure faster convergence towards the optimal solution.

\section{Related Work}
\label{sec:Related Work}
Earlier work on application level tuning of transfer parameters mostly proposed static or non-scalable solutions to the problem with some predefined values for some generic cases~\cite{globusonline, NDM_2012, R_Hacker02, R_Crowcroft98, R_Dinda05}. The main problem with such solutions is that they do not consider the dynamic nature of the network links and the background traffic in the intermediate nodes. Managed File Transfer (MFT) systems were proposed which used a subset of these parameters in an effort to improve the end-to-end data transfer throughput~\cite{WORLDS_2004, ScienceCloud_2013, globusonline, Royal_2011, IGI_2012}.

Yin et al. ~\cite{R_Yin11} proposed a full second order model with at least three real-time sample transfers to find optimal parallelism level. The relationship between parallel streams and throughput along with other parameters are more complex than second order polynomials. Moreover, it does not provide concurrency and pipelining. Yildirim et al.~\cite{R_Yildirim11}, Kim et al.~\cite{DISCS12} proposed similar approaches with higher order polynomials. 
 
Yildirim et al.~\cite{balance} proposed PCP algorithm which clusters the data based on file size and performs sample transfers for each cluster. Sampling overhead could be very high in this model as it does not consider any historical knowledge for optimization. 

Engin et al. ~\cite{Engin2016} proposed HARP which uses heuristics to provide initial transfer parameters to collect data about sample transfers. After that model performs the optimization on the fly where it has to perform cosine similarity over the whole dataset which might prove expensive. Even if the optimization and transfer task can be parallelized, it could be wasteful as the same optimization needs to be performed for similar transfers every time a similar transfer request is made.  
 
 Prasanna et al. ~\cite{Ian-bala2016} proposed direct search optimization that tune parameters on the fly based on measured throughput for each transferred chunk. However, it is hard to prove the convergence and sometimes hard to predict the rate of convergence. Some cases, it requires 16-20 epochs to converge which could lead to under-utilization. 
Liu et al. ~\cite{Liu:2017} explored Globus historical logs consisting of millions of transfers to analyze the effects of tunable parameters on the transfer characteristics. 
 
 Different from the existing work, we address the following issues in this paper:
 \begin{enumerate}
  \item Lower order regression model can underfit the data when higher order polynomials can introduce overfitting, in addition, to compute cost and sampling overhead. For small to moderate size of data transfer requests, slow convergence could lead to severe under-utilization.
  
  \item Model free dynamic approaches suffer from convergence issue. And convergence time depends on the location of initial search point.
  
  \item Searching parameters during the transfer could introduce many overheads. Opening a TCP connection in the middle of the transfer introduces a delay due to slow start phase. When initial parameters are far away from optimal solution slow convergence could lead to under-utilization of the network bandwidth which could hurt the overall bandwidth.  
 
 \item  Optimization based on historical log should not be done during the transfer, offline analysis can reduce the real-time computing overhead. 
\end{enumerate}

\section{Conclusion}
\label{sec:conclusion}
In this study, we have explored a novel data transfer throughput optimization model that relies upon offline mathematical modeling and online adaptive sampling. Existing literature contains different types of throughput optimization models that range from static parameter based systems to dynamic probing based solutions. Our model eliminates online optimization cost by performing the offline analysis which can be done periodically. It also provides accurate modeling of throughput which helps the online phase to reach near optimal solution very quickly. For large scale transfers when external background traffic can change during transfer, our model can detect the harsh changes and can act accordingly. Adaptive sampling module can converge faster than existing solutions. The overall model is resilient to harsh network traffic changes. We performed extensive experimentations and compared our results with best known existing solutions. Our model outperforms existing models in terms of accuracy, convergence speed, and achieved throughput. The throughput prediction accuracy of our model reaches 93\%. 

As future work, we are planning to increase the achievable throughput further by reducing the impact of TCP slow start phase. Another interesting path is to reduce the overhead introduced by real-time parameter changes. We are also planning to investigate other application-layer protocol parameter sets that can be optimized to achieve even better performance.

\bibliographystyle{plain}
\bibliography{main,didc,misc,rela_work} 

\end{document}